\documentclass[11pt]{article}
\usepackage{amsfonts,amssymb}
\usepackage{graphicx}
\usepackage[totalwidth=16cm,totalheight=23.5cm]{geometry}
\usepackage{epigraph} 

\newcommand{\C}{{\mathbb C}}
\newcommand{\R}{{\mathbb R}}

\newcommand{\im}{{\rm i }}

\newcommand\be{\begin{eqnarray}}
\newcommand\ee{\end{eqnarray}}

\begin{document} 

\pagenumbering{roman}

\begin{titlepage}

\title{A Gauge Theoretic Approach to Gravity} 
\author{Kirill Krasnov \\ \\
{\it School of Mathematical Sciences, University of Nottingham}\\ {\it University Park, Nottingham, NG7 2RD, UK} \\ and \\
{\it Max Planck Institute for Gravitational Physics (Albert Einstein Institute)} \\ {\it Am M\"uhlenberg 1, 14476 Golm, Germany}}
\date{February 2012} 
\maketitle

\begin{abstract}Einstein's General Relativity (GR) is a dynamical theory of the spacetime metric. We describe an approach in which GR becomes an ${\rm SU}(2)$ gauge theory. We start at the linearised level and show how a gauge theoretic Lagrangian for non-interacting massless spin two particles (gravitons) takes a much more simple and compact form than in the standard metric description. Moreover, in contrast to the GR situation, the gauge theory Lagrangian is convex. We then proceed with a formulation of the full non-linear theory. The equivalence to the metric-based GR holds only at the level of solutions of the field equations, that is, on-shell. The gauge-theoretic approach also makes it clear that GR is not the only interacting theory of massless spin two particles, in spite of the GR uniqueness theorems available in the metric description. Thus, there is an inifnite-parameter class of gravity theories all describing just two propagating polarisations of the graviton. We describe how matter can be coupled to gravity in this formulation and, in particular, how both the gravity and Yang-Mills arise as sectors of a general diffeomorphism invariant gauge theory. We finish by outlining a possible scenario of the UV completion of quantum gravity within this approach. 
\end{abstract}

\end{titlepage}

\pagenumbering{arabic}

\epigraph{Since the mathematicians have invaded the theory of relativity, I do not understand it myself anymore.}%
{\textit{Albert Einstein}}

\section{Introduction} 

This expository paper is about an approach to the theory of general relativity (GR) that departs rather far from how its initiator Albert Einstein thought about the subject. The present author is a physicist by training but working in a mathematics department, and so it is quite possible that Einstein would be able to say about this work what he said on a different occasion and what is quoted above. The approach we shall describe makes a conceptual jump as compared to Einstein's views on the nature of the gravitational force. Thus, in Einstein's general relativity the gravitational field is encoded in the geometrical properties of the spacetime, and the geometry itself is described by the spacetime metric. The approach described here will still be  geometrical in spirit. However, the geometry will be that of {\it gauge fields} on the spacetime manifold, not that of a spacetime metric. No metric will be present in the formulation of the theory, and it will only appear later as a derived concept. 

Accepting for the moment  that such an approach is indeed possible, a natural question to ask is if it is worth considering any such reformulation of GR. Indeed, it is clear that it would require drastic changes to the century old metric-based intuition about the subject. Such conceptual changes can only be justified if they can help to solve problems that are difficult to address in the traditional framework. As we now briefly review, there are indeed at least two major difficulties with General Relativity. 

The first problem of GR is that of the unification with the other forces. Thus, Einstein's theory was developed already almost 100 years ago, in particular, before the theory of quantum mechanics was discovered. And during the last century our understanding of the behaviour of the micro-world has culminated in the discovery that almost all interactions in Nature (apart from gravity) are described by (quantum) gauge fields. Thus, it is now believed that (to the energy scales of the conjectural grand unification) the electromagnetic, the weak and the strong interactions are all described by the Yang-Mills theory of the (spontaneously broken) Standard Model gauge group ${\rm U}(1)\times {\rm SU}(2)\times{\rm SU}(3)$. In other words, at least to a very high energy scale the gauge fields appear to be fundamental. Only the gravitational force stands far aside from this compelling pattern. Using the technical jargon we say that gauge fields describe particles of spin 1, while the gravitational interactions are carried by particles of spin 2, and this is where the difference comes from. At the same time, if it was possible to reformulate the theory of gravity in the language closer to that of gauge theories, we would possibly come closer to understanding how gravity fits together with the other forces of Nature. To put it differently, it is not clear that Einstein himself would be thinking in terms of the spacetime metric if he was trying to invent his theory of gravity nowadays, with our present solid appreciation of the importance of the gauge principle. 

The other problem of GR is related to the quantum properties of the gravitational field. These do not seem to be of any importance at energy scales accessible to our observations or experiments. At the same time, they can be studied theoretically and reveal that gravity is again unlike all the other forces. For other forces the quantum vacuum polarisation effects lead to the strength of interactions becoming energy-scale dependent (the phenomenon of renormalisation). Yet, the form of the field equations (or the Lagrangian) describing these interactions does not change with the energy, only the coupling constants (and other physical parameters e.g. masses) flow with the scale. One says that the theories that form the Standard Model of elementary particles are renormalisable. Gravity is not like this. Not only the parameters of its Lagrangian flow with the energy, but also the form of the Lagrangian itself must be modified once the quantum corrections are taken into account. One says that gravity is non-renormalisable. What this means is that we know for certain that Einstein's theory of general relativity cannot be a fundamental theory, for at very high energies it is modified (by quantum corrections) in a way we cannot currently control. 

A possible way out of the above mentioned difficulties is to imbed both gravity and the gauge-fields in a different theory that resolves these problems. This is e.g. what string theory attempts to do. However, the string theory, at least in the way we currently understand it, appears to be far from being an economical solution of the problem, as it requires supersymmetry as well as many extra unseen dimensions. It is not impossible that a more economical solution to the problems of gravity could exist. 

As we shall explain in this article, the present gauge theory approach may be of help with both the problem of unification with other interactions as well as the problem of the quantum behaviour. We emphasise, however, that none of these two goals has been achieved as of yet, there are only hints (to be described) that this may be possible. A sceptically minded reader may then conclude that in the absence (as of yet) of any real applications the approach presented here is not worth the trouble learning, for it requires  developing a completely new intuition (gauge-field based) about the gravitational force.  

At the same time we note that in the past the science has often progressed by first reformulating a known physical law in a non-trivially equivalent form, and then this non-trivial reformulation suggesting a generalisation leading to a new physical theory. For this author, the most striking example of this historical phenomenon is the discovery of quantum mechanics. Indeed, at least in its first formulations it was so essentially built on the Hamiltonian formulation of the classical physics. Thus, it appears to be profitable to reformulate known physical laws in a form that is non-trivially equivalent to the original formulation. This way we not only learn more about the theory at hand, but also potentially step on the path to new physics. We therefore proceed with our development of a gauge-theoretic description of gravity having at least this historic lesson as the motivation.

As for a mathematically minded reader, we motivate our constructions by the following remarks. In spite of Yang-Mills theory and gravity being so different in spirit --- the former is about connections in principal G-bundles while the later is about metric-compatible connections in the tangent bundle --- a subtle interplay between the two theories has become apparent via work on Yang-Mills instantons in late 70's. This is particularly clearly illustrated by the fact that the index 1 ${\rm SU}(2)$ instantons (self-dual solutions of the Yang-Mills field equations) are all gauge-equivalent to the self-dual part of the gravitational Levi-Civita connection for metrics conformally related to the standard round metric on $S^4$, see Theorem 9.1 in \cite{Atiyah:1978wi}. Rephrasing, one can say that all index 1 ${\rm SU}(2)$ instantons come from gravitational instantons. Thus, at least in the self-dual case, there is an intimate relation between the solutions of the two theories. It is this relation that is going to be at the root of the developments reviewed in this paper, with the notion of the self-duality playing the central role. 

The final word of caution is as follows. We make no claim here that by simply reformulating gravity in the language of gauge fields one comes closer to understanding the problems of quantum gravity and unification. In fact, such reformulations were around for a long time, with historically the first being those due to Plebanski \cite{Plebanski:1977zz} and MacDowell-Mansouri \cite{MacDowell:1977jt}, then followed by a new Hamiltonian formulation due to Ashtekar \cite{Ashtekar:1987gu}. These formulations of GR do shed some new light on the problems of gravity, but do not by themselves provide a solution of these problems. So, the novelty of the approach described in this paper is not in the fact that a gauge-theoretic reformulation of gravity is possible --- this was known for quite some time. It is in the fact that the specific gauge theory description to be reviewed here has certain attractive features not shared by other formulations, and these may eventually help us to come to terms with the above mentioned problems of gravity. Thus, one of the most interesting features of the new formulation is that the (Euclidean signature) action functional becomes convex, a very desirable property that the usual metric-based formulation does not have. This feature is in turn related to other simplifications in the structure of the theory, which is what gives us hope that the new formulation can shed new light on the problems of gravity.  

With these introductory remarks being made, we shall proceed with our gauge-theoretic description of gravity. We shall follow the bottom-to-top approach, in which we will first present a gauge theory description of gravity in as least general terms as possible, and then proceed to exhibit the general principle. Thus, in Section \ref{sec:lin} we show how the linearized theory (i.e. gravitons) can be described in the gauge theory terms. We present the construction of the full non-linear theory in Section \ref{sec:digt}. This section also discussed how the unification is realised in the present approach. We then discuss the quantum theory in Section \ref{sec:quant}. We reiterate the main points of our approach in Section \ref{sec:concl}. 

\section{Gravity as a gauge theory - linearised level}
\label{sec:lin}

We start by showing how gravity can be described in gauge-theoretic terms at the level of the linearised theory. This is already non-trivial, because the statement that gauge fields are spin one, while gravitons are spin two particles is the statement about the corresponding linearised theories. 

From the outset we note that the fact that the linearized general relativity can be described in terms of connections is not new. For instance, this is possible in the framework of Plebanski formulation of GR \cite{Plebanski:1977zz} and related to it (via the space plus time decomposition) Ashtekar Hamiltonian formulation \cite{Ashtekar:1987gu}. The principal difference between these and our approach is that our linearized theory Lagrangian will be a functional of only the connection field, with no other (auxiliary) fields necessary (as in e.g. \cite{Plebanski:1977zz}). We shall see that this leads to some important simplifications, and that our linearized theory will be arguably simpler than all other known descriptions of the linearized gravity. In particular, as we shall see below, the diffeomorphisms are realized in our gauge-theoretic description in a particularly simple way (different from all other existing descriptions), and the corresponding gauge components of the field can be projected away from the very beginning, leading to considerable simplifications in the structure of the theory. In addition, we shall see that our Lagrangian for the linearised GR has very much the same form as that for the linearised Yang-Mills theory, which will eventually be of help when we put the two theories together. 

The gauge-theretic description of free massless spin 2 particles given in this section appears to be new, and was first given in \cite{Krasnov:2011up}.

\subsection{Linearised Yang-Mills theory}

Let us start our description by briefly reviewing the linearization of the Yang-Mills theory. Let $A$ be a connection in the principal G-bundle over the spacetime $M$, and let $F=dA+(1/2)[A,A]$ be its curvature two-form. Here we assume $G$ to be compact. We use the conventions in which $F=(1/2)F_{\mu\nu} dx^\mu\wedge dx^\nu$ and often suppress the Lie-algebra index of $F$ for brevity. If $g_{\mu\nu}$ is a (given) spacetime metric, which we for now assume to be that of the Minkowski spacetime, then it can be used to raise the indices of $F_{\mu\nu}$ to get $F^{\mu\nu}$, which can then be contracted with $F_{\mu\nu}$ to form the familiar Yang-Mills Lagrangian 
\be\label{L-YM}
{\cal L}_{\rm YM}=-\frac{1}{4 g_{YM}^2} {\rm Tr} (F^{\mu\nu} F_{\mu\nu}),
\ee
where the trace stands for a (in our conventions positive-definite) Killing-Cartan form on the Lie algebra of $G$, appropriately normalised. Thus, the trace can also be replaced by a sum over the Lie-algebra indices of the curvature squared. At the linearised level (expanding around the zero background connection and introducing the canonically normalised perturbation $a$ so that  $A=g_{YM} \, a$) we get
\be
{\cal L}_{\rm YM}^{(2)} = -\frac{1}{ 2} {\rm Tr} \left(\partial^{\mu} a^{\nu} (\partial_\mu a_\nu-\partial_\nu a_\mu) \right).
\ee
As usual, the need for the minus sign in front of the action is dictated by the desire to have a positive-definite Hamiltonian. Indeed, introducing the space+time split $\mu=(0,i), i=1,2,3$ and taking (for definiteness) the metric signature to be $(-,+,+,+)$ we get
\be
{\cal L}_{\rm YM}^{(2)} = \frac{1}{ 2} {\rm Tr} \left( (\dot{a}_i-\partial_i a_0)^2 - (B_i)^2 \right),
\ee
where the dot denotes the time derivative and we have introduced a notation $B_i = \epsilon_{ijk} \partial_j a_k$. We therefore see that the momentum canonically conjugate to $a_i$ is $E_i := \dot{a}_i - \partial_i a_0$, and the the momentum conjugate to the time component of the connection $a_0$ is zero. Thus, the time component of the connection plays the role of the Lagrange multiplier imposing the so-called Gauss constraint, and the linearised Hamiltonian ${\cal H}_{\rm YM}^{(2)}= {\rm Tr}(E_i \dot{a}_i) - {\cal L}_{\rm YM}^{(2)} $ is given by:
\be\label{H-YM}
{\cal H}_{\rm YM}^{(2)} = \frac{1}{ 2} {\rm Tr} \left( (E_i)^2 +(B_i)^2 \right) - {\rm Tr}(a_0 \partial_i E_i),
\ee
where we have integrated by parts in the last term. The Hamiltonian is explicitly positive-definite (if the gauge group $G$ is compact), modulo a constraint term that imposes the electric field to be tracefree $\partial_i E_i=0$ and generates the usual gauge transformations of the connection $\delta_\phi a_i = \partial_i \phi$, $\phi$ being an arbitrary Lie-algebra-valued function. The Hamiltonian is easily seen to be gauge-invariant, as the electric field does not transform (at this linearised level) and the linearised connection transforms so that the magnetic field $B_i$ is gauge-invariant. 

\subsection{Self-duality}

In preparation to the developments in the case of gravity, we will add to the Lagrangian (\ref{L-YM}) a certain total derivative term. This will have no effect on the physics, but will allow us to write the YM Lagrangian in a form later almost matched by the gravitational one. To write down the total derivative term required we need to introduce some new object that will play a very important role later. These objects are certain self-dual two-forms and thus build on the notion of the self-duality on two-forms (in 4 spacetime dimensions that we work in). 

Recall that, given a metric $g_{\mu\nu}$ (such as e.g. the Minkowski metric we have been working with up to now) one has the volume form $\epsilon_{\mu\nu\rho\sigma}$ for $g_{\mu\nu}$. Two of its indices can be raised with the metric to obtain an object $\epsilon_{\mu\nu}{}^{\rho\sigma}$ that can be applied to a two-form $U_{\mu\nu}$ with the result being again a two-form -- the Hodge dual of $U$. We have:
\be
U^*_{\mu\nu}:=\frac{1}{2} \epsilon_{\mu\nu}{}^{\rho\sigma} U_{\rho\sigma}.
\ee
As is not hard to check, in our case of a metric of the Lorentzian signature the square of the Hodge operator is minus one. Thus, its eigenvalues are $\pm \im$, and the space of two-forms splits into two orthogonal subspaces. These are referred to as the spaces of self-dual and anti-self-dual two-forms. We note in passing that these properties of the Hodge duality operator are quire reminiscent of those of a complex structure (on the space of two-forms), with the spaces of self- and anti-self-dual forms being analogous to holomorphic and anti-holomorphic elements. This is more than analogy, and we refer the interested reader to \cite{Brans:1974ry} for a development of this idea. 

Thus, for any self-dual two-form $U_{\mu\nu}$ we have:
\be
\frac{1}{2} \epsilon_{\mu\nu}{}^{\rho\sigma} U_{\rho\sigma} = \im U_{\mu\nu},
\ee
and for anti-self-dual form we have an extra minus on the right-hand-side. The space of self-dual two-forms is 3-dimensional, and we can introduce a basis in it. A choice of such basic self-dual two-forms can be rather arbitrary as long as they span the required subspace. However, there is always a canonical (modulo certain gauge rotations, see below) {\it orthonormal} choice of the basis. Let us denote the canonical self-dual two-forms by $\Sigma^i_{\mu\nu}, i=1,2,3$. Note that we have denoted the index enumerating the two-forms by the same letter as was used to refer to the spatial index in the Hamiltonian analysis above. This is not an oversight; we shall now see that the two indices can be naturally identified. The canonical self-dual two-forms are required to be orthonormal in that 
\be\label{metricity}
\epsilon^{\mu\nu\rho\sigma} \Sigma^i_{\mu\nu}  \Sigma^j_{\rho\sigma}  = 8\im  \delta^{ij},
\ee
where the numerical coefficient on the right-hand-side is convention-dependent, and $\delta^{ij}$ is the Kronecker-delta. It can be shown that the self-dual two-forms satisfying (\ref{metricity}) are defined uniquely modulo ${\rm SO}(3)$ rotations preserving $\delta^{ij}$. We can now give an explicit form of the basic self-dual two-forms in the case of the Minkowski spacetime metric. Using the two-form notation we have:
\be\label{Sigma}
\Sigma^i=\im dt\wedge dx^i + \frac{1}{2} \epsilon^{ijk} dx^j\wedge dx^k.
\ee
it is not hard to check the $\Sigma^i_{\mu\nu}$ are self-dual (with the conventions that $\epsilon^{0123}=+1$), and that (\ref{metricity}) holds. Let us also note what becomes of the components of the two-forms $\Sigma^i_{\mu\nu}$ under the space+time split. We have:
\be
\Sigma^i_{0j} = \im \, \delta^i_j, \qquad \Sigma^i_{jk} = \epsilon^{i}{}_{jk}.
\ee
Thus, we see that the objects $\Sigma^i_{\mu\nu}$ indeed provide a natural identification of the basis index $i$ with the spatial index. This fact will be used later in our discussion of the gravitational case.

For later purposes, let us note an important identity satisfied by our self-dual two-forms. We have
\be\label{S-algebra}
\Sigma^i_\mu{}^\nu \Sigma^j_\nu{}^\rho = -\delta^{ij} g_\mu{}^\rho +\epsilon^{ijk} \Sigma^k_\mu{}^\rho.
\ee
Thus, the basic self-dual two-forms satisfy an algebra similar to that of Pauli matrices. This identity can be checked by a direct verification, using the explicit expression (\ref{Sigma}).

For our discussion of the Yang-Mills theory we need one more identity involving $\Sigma^i_{\mu\nu}$. Thus, we note that we can use $\Sigma$'s to construct the projector on the space of self-dual two forms. This is an operator $P^+$ whose square coincides with itself, and which projects any two-form onto its self-dual part. We have:
\be
P_{\mu\nu}^{+\quad\rho\sigma} \equiv \frac{1}{2} \left( g_{[\mu}^\rho g_{\nu]}^\sigma - \frac{\im}{2} \epsilon_{\mu\nu}{}^{\rho\sigma} \right) = \frac{1}{4} \Sigma^i_{\mu\nu} \Sigma^{i\,\rho\sigma}.
\ee

Finally, we note that the self-dual two-forms $\Sigma^i$ have appeared in the literature on many occasions before. In the literature on instantons (self-dual solutions of the YM field equations) these objects are often referred to as 't Hooft's symbols \cite{'tHooft:1976fv}. They will be of fundamental importance in the considerations that follow.

\subsection{Linearised Yang-Mills revisited}

We now use the objects introduced in the previous subsection to rewrite the linearised YM Lagrangian (plus a surface term) in a convenient for the later purposes form. Thus, modulo a total derivative term we have:
\be
{\cal L}_{\rm YM}^{(2)} = -2 P^{+\,\mu\nu\rho\sigma} {\rm Tr}(\partial_\mu a_\nu \partial_\rho a_\sigma).
\ee
Using the basic self-dual two-forms $\Sigma^i_{\mu\nu}$ introduced above we can rewrite this as
\be
{\cal L}_{\rm YM}^{(2)} = -\frac{1}{2} {\rm Tr}( \Sigma^{i\,\mu\nu}\partial_\mu a_\nu)^2.
\ee
An alternative convenient way to write the above Lagrangian is to introduce a basis in the Lie algebra, so that $a_\mu = a^a_\mu T^a$, with $a$ being the Lie-algebra index and $T^a$ being the generators normalised so that ${\rm Tr}(T^a T^b)=\delta^{ab}$. With this in mind we rewrite the YM Lagrangian as
\be\label{L-YM-S}
{\cal L}_{\rm YM}^{(2)} = -\frac{1}{2} \delta_{ab} \delta_{ij} ( \Sigma^{i\,\mu\nu}\partial_\mu a^a_\nu)( \Sigma^{j\,\rho\sigma}\partial_\rho a^b_\sigma).
\ee
Our linearised gravitational Lagrangian below will strongly resemble this form of the YM Lagrangian. 

An instructive exercise is to repeat the Hamiltonian analysis of YM but starting from (\ref{L-YM-S}). To this end, we need the space+time split of the combination $ \Sigma^{i\,\mu\nu}\partial_\mu a^a_\nu$ that appears prominently in (\ref{L-YM-S}). We have:
\be\label{S-da}
\Sigma^{i\,\mu\nu}\partial_\mu a^a_\nu = - \im (\dot{a}^a_i - \partial_i a_0^a ) +\epsilon^{ijk} \partial_j a_k^a.
\ee
We now see that in this formulation of the theory, the momentum conjugate to $a_i$ is 
\be\label{pi-YM}
\pi_i:=\dot{a}_i -\partial_i a_0 +\im B_i = E_i + \im B_i,
\ee
where $E_i$ is the usual electric field and we have used the same as above definition of the magnetic field $B_i$. We have again suppressed the Lie-algebra index for brevity. The Hamiltonian in this formulation reads
\be
{\cal H}_{\rm YM}^{(2)} = {\rm Tr}\left(  \frac{1}{2}( \pi_i)^2 - \im\, \pi_i B_i \right) - {\rm Tr}(a_0 \partial_i \pi_i).
\ee
This can be rewritten as
\be\label{H-YM-im}
{\cal H}_{\rm YM}^{(2)} =\frac{1}{2} {\rm Tr}\left(  ( \pi_i-\im B_i)^2 + ( B_i)^2 \right) - {\rm Tr}(a_0 \partial_i \pi_i),
\ee
which is the same Hamiltonian as we have obtained above in (\ref{H-YM}). Indeed, for the first term this is obvious, and for the last Gauss constraint term the shift of the momentum $\pi_i$ by a multiple of $B_i$ has no effect as the magnetic field is transverse. We see that in this formulation the momentum conjugate to the connection is not real, but rather only the combination $\pi_i -\im B_i\equiv E_i$ is. At the same time the spatial part $a_i$ of the connection is real.  

The reason why the complexity crept in is that we have added to the Lagrangian a term that is a {\it purely imaginary} total derivative. This has no effect on the dynamics, but it does affect the structure of the Hamiltonian formulation of the theory, in particular the symplectic structure. This is why the momentum variable is no longer the electric field, but rather the combination (\ref{pi-YM}). This corresponds to a canonical transformation on the phase space of YM theory with a purely imaginary generating function. The analysis of this subsection is of course very familiar from the discussions of the $\theta$-angle ambiguity in the quantisation of YM theory, see e.g. Section 13.2 of \cite{Ashtekar:1991hf} for a very readable account. The only difference with the more familiar case is that the parameter of the canonical transformation is taken to be imaginary. This is certainly a legitimate operation at the classical level, with the price to pay being that one has to be careful about the reality conditions satisfied by the fields. We shall see that similar care will have to be exercised in the gravitational case, where analogous reality conditions are to be imposed.

\subsection{Linearised gravity}

We now turn to the main object of our interest -- gravity. We start by simply writing down the linearised Lagrangian of a form similar to (\ref{L-YM-S}), and later explain how it comes about. 

The first step is a choice of the gauge group, and we take $G={\rm SO}(3)$. We shall then denote the Lie algebra index by the same lower case latin letters from the middle of the alphabet, which we have already used to refer to the spatial indices. Again this is not an oversight, since we shall later see that these types of indices can be naturally identified. We then write a Lagrangian essentially of the same form as (\ref{L-YM-S}), but with a different tensor used to contract the two quantities $ \Sigma^{i\,\mu\nu}\partial_\mu a^a_\nu$, where $a$ is now to be replaced by an ${\rm SO}(3)$ index $j$. Thus, we write
\be\label{L-GR}
{\cal L}_{\rm GR}^{(2)} = -\frac{1}{2} P^{ij|kl} ( \Sigma^{i\,\mu\nu}\partial_\mu a^j_\nu)( \Sigma^{k\,\rho\sigma}\partial_\rho a^l_\sigma),
\ee
where
\be\label{P}
P^{ij|kl}:=\delta^{i(k} \delta^{l)j} - \frac{1}{3}\delta^{ij}\delta^{kl}
\ee
is a projector on symmetric tracefree $3\times 3$ matrices. As we shall soon see, the Lagrangian (\ref{L-GR}) describes a spin two particle, in contrast to (\ref{L-YM-S}) that describes a spin one particle for each Lie algebra generator. The fact that (\ref{L-GR}) gives a spin two particle is related to the fact that (\ref{P}) is a projector on spin two representation in the tensor product of two spin one representations of ${\rm SO}(3)$. 

Let us now discuss the invariances of the Lagrangian (\ref{L-GR}). As in the case of Yang-Mills theory, (\ref{L-GR}) is obvious invariant under the ${\rm SO}(3)$ gauge-transformations $\delta_\phi a^i_\mu = \partial_\mu \phi^i$. Indeed, the partial derivatives commute, and the objects $\Sigma^{i\,\mu\nu}$ are anti-symmetric in their spacetime indices. However, unlike in the YM case, the Lagrangian (\ref{L-GR}) has an extra invariance, and this is how the diffeomorphisms appear in the game. Consider
\be\label{diffeo}
\delta_\xi a^i_\mu = \xi^\nu \Sigma^i_{\mu\nu},
\ee
where $\Sigma^i_{\mu\nu}$ are the basic self-dual two-forms (\ref{Sigma}), and $\xi^\mu$ is an arbitrary vector field. Note that there are no derivatives present in this transformation rule, this is simply a shift of the linearised connection in a certain direction in the field space. It is not hard to see that our Lagrangian is invariant under such shifts. Indeed, we use the algebra (\ref{S-algebra}) of the basic self-dual two-forms and note that the product of two $\Sigma$'s is in either the spin zero part (the trace) or in the spin one part (the anti-symmetric part) in the tensor product of two spin one representations. However, the Lagrangian contains the spin two projector (\ref{P}), and so is insensitive to the shifts (\ref{diffeo}). So, (\ref{diffeo}) is an invariance of the Lagrangian. However, as we shall soon see, this is not a gauge invariance in the sense how the usual gauge transformations of the connection were the gauge invariances of the theory in the case of Yang-Mills theory. Thus, we shall see that there are no Lagrange multipliers for the constraints generating (\ref{diffeo}) in the Hamiltonian formulation of the theory. The significance of this fact will be discussed later.

Let us now repeat the exercise of the Hamiltonian analysis of the theory. We have already made most of the steps needed for this in the discussion of the YM theory. So, we can simply write the expression for $\Sigma_i^{\mu\nu}\partial_\mu a_{\nu j}$ from (\ref{S-da}). We have:
\be
\Sigma_i^{\mu\nu}\partial_\mu a_{\nu j} = - \im (\dot{a}_{ij} - \partial_i a_{0j} ) +\epsilon_i{}^{kl} \partial_k a_{lj}
\ee
where the second index on the spatial connection is the lowered Lie algebra one. In the Lagrangian this expression is multiplied by the spin two projector (\ref{P}). Thus, we immediately see that only the spin two part of the connection field $a_{ij}$ is dynamical. The momentum conjugate to this part of the connection is 
\be\label{pi-grav}
\pi^{ij} = P^{ijkl} \left((\dot{a}_{kl} - \partial_k a_{0 l} ) +\im \, \epsilon_k{}^{mn} \partial_m a_{nl} \right).
\ee
We see that this expression depends only on the time derivative of the symmetric tracefree part of $a_{ij}$, on the temporal component of the connection $a_{0i}$, as well as on spatial derivatives of the full connection $a_{ij}$. Let us now discuss the dependence on the latter. For this, it is very convenient to split the spatial connection $a_{ij}$ into its irreducible with respect to ${\rm SO}(3)$ components. Thus, we have:
\be
a_{ij} = a^{(2)}_{ij} + a^{(1)}_{ij} + a^{(0)}_{ij},
\ee
where $a^{(2)}_{ij}$ is the spin two part that is symmetric tracefree, $a^{(1)}_{ij}$ is the anti-symmetric spin one part, and $a^{(0)}_{ij}$ is the spin zero part proportional to $\delta_{ij}$. We immediately see that because of the symmetrization imposed by the projector $P^{ij|kl}$ there is no dependence in the last term in (\ref{pi-grav}) on the spin zero part $a^{(0)}_{ij}$. Next, let us write 
\be
a^{(1)}_{ij} = \epsilon_{ijk} c^k,
\ee
where $c^k$ captures the spin one part of the spatial connection. Then a simple calculation shows that the momentum (\ref{pi-grav}) can be written as 
\be\label{pi-grav-1}
\pi^{ij} = P^{ijkl} \left(\dot{a}^{(2)}_{kl} +\im \, \epsilon_k{}^{mn} \partial_m a^{(2)}_{nl} - \partial_k (a_{0 l}+\im c_l)   \right).
\ee
We thus see that the Lagrangian (\ref{L-GR}) only depends on the spin two part of the spatial connection $a_{ij}$, as well as on the temporal $a_{0i}$ and spin one components $a^{(1)}_{ij}$ in the combination $a_{0 i}+(\im/2)\epsilon_i{}^{jk} a^{(1)}_{jk}$. There is no dependence at all on the spin zero part $a^{(0)}_{ij}$.

Let us now find the Hamiltonian ${\cal H}^{(2)}_{\rm GR}=\pi^{ij} a^{(2)}_{ij} - {\cal L}^{(2)}_{\rm GR}$. We have:
\be
{\cal H}^{(2)}_{\rm GR} = \frac{1}{2}(\pi^{ij})^2 -\im \pi^{ij} \epsilon_i{}^{kl} \partial_k a_{lj} - (a_{0i}+\im c_i) \partial_j \pi^{ij}.
\ee
Here we have removed the superscript $(2)$ from the spin two component of the connection for brevity, with the understanding now that this is the only component of the connection that is dynamical. We have also removed all the projectors $P^{ij|kl}$ because now the momentum $\pi^{ij}$ is symmetric traceless and so no additional projection is necessary. In addition, we have integrated by parts in the last term.

We see the structure already very familiar from the discussion of the Yang-Mills theory above. We could write the Hamiltonian in a more suggestive form by introducing the "magnetic field" 
\be
B^{ij}:=P^{ij|kl} \epsilon_k{}^{mn} \partial_m a_{nl}
\ee
and writing
\be\label{H-GR}
{\cal H}^{(2)}_{\rm GR} = \frac{1}{2}\left( (\pi^{ij}-\im B^{ij})^2 + (B^{ij})^2 \right) - (a_{0i}+\im c_i) \partial_j \pi^{ij}.
\ee
This is exactly analogous to (\ref{H-YM-im}), with the only exception being that the momentum variable $\pi^{ij}$ as well as the magnetic field $B^{ij}$ are now spin two fields. As in the YM case, the last term, when varied with respect to the combination $a_{0i} +\im c_i$, generates the constraint that requires the momentum to be transverse. The reality conditions that need to be imposed to guarantee the positive-definiteness of the Hamiltonian are also analogous to the YM case. Indeed, we see that the combination $\pi^{ij}-\im B^{ij}$ must be required to be real, as well as the spin two part $a^{(2)}_{ij}$ of the spatial connection (and thus the magnetic field $B^{ij}$). 

Thus, we see that at the level of the Hamiltonian formulation (as well as at the level of the action) the treatments of Yang-Mills theory and gravity are exactly analogous. The main difference comes from the fact that in the gravity case the gauge group has to be taken a specific one ${\rm SO}(3)$, and that certain propagating modes of the Yang-Mills theory with its two modes per generator have been projected out to leave only two propagating degrees of freedom of gravity. Indeed, the effect of the projector $P^{ij|kl}$ inserted in the Lagrangian (\ref{L-GR}) is precisely to remove 4 of the 6 propagating modes of ${\rm SO}(3)$ YM theory, leaving only two gravitational DOF. At the level of the Hamiltonian (\ref{H-GR}) this projection is already carried out because only the spin 2 part of the spatial connection (and its conjugate momentum) are present. 
 
The above discussion of the propagating mode content of (\ref{L-GR}) can be rephrased by saying that in the case of YM theory we have three modes per each group generator together with one constraint, and thus two propagating modes per generator. They are described by the spatial connection $a_i$ (with the Lie algebra index suppressed), which can be set (using gauge transformations) to satisfy the transverse condition $\partial^i a_i =0$. In the case of gravity we have exactly the same starting point, and the spatial connection $a_i^j$, where now the Lie algebra index is explicitly indicated. Since the group ${\rm SO}(3)$ of gauge rotations in this case can be identified with the group of spatial rotations (this identification is provided e.g. by the spatial projection of the self-dual two-forms $\Sigma^i_{\mu\nu}$), the two indices of $a_i^j$ are on the same footing. Thus, the spatial connection in this case is in the product of two spin one representations: $a\in V^1\otimes V^1$, where our notation is that $V^j$ is the space of irreducible representation of ${\rm SO}(3)$ of dimension ${\rm dim}(V^j)=2j+1$. Decomposing $V^1\otimes V^1$ into irreducibles we get a direct sum $V^2\oplus V^1\oplus V^0$. The projector in (\ref{L-GR}) projects out the spin one and spin zero components, leaving only a single dynamical spin two field $a^{(2)}_{ij}$. Gauge transformations (${\rm SO}(3)$ rotations) still act on it and can be used to set this field to be transverse. Once this is done the term $(B^{ij})^2$ becomes $(\partial_i a_{jk})^2$, as can be checked by a simple calculation. Thus, once the gauge has been fixed we get the following linearised Hamiltonian:
\be
{\cal H}^{(2)}_{\rm GR} = \frac{1}{2}\left( (p^{ij})^2 + (\partial_i a_{jk})^2 \right),
\ee
where we have denoted the real momentum by $p^{ij}=\pi^{ij}-\im B^{ij}$. This is the usual linearised Hamiltonian of the metric-based GR, as we shall review in the next subsection.

\subsection{A comparison to the linearised Einstein-Hilbert action}

Above we have seen how a spin two particle can be described by a Lagrangian of the YM type, but with an additional spin two projector inserted. At the level of the Hamiltonian formulation we have seen the familiar from the metric-based treatment symmetric traceless transverse tensors appearing. The natural question that arises is how the above Lagrangian is related to the usual metric-based GR Lagrangian linearised around the Minkowski metric. The purpose of this subsection is to discuss this relationship.  

The Einstein-Hilbert Lagrangian (for our choice of the signature (-,+,+,+)) reads:
\be\label{L-EH}
{\cal L}_{\rm EH} = -\frac{2}{\kappa^2} \sqrt{-g} R,
\ee
where $\kappa^2=32\pi G$, with $G$ being the Newton's constant, and $R$ is the Ricci scalar. We take the background metric to be the Minkowski one, and linearise (\ref{L-EH}) taking the perturbation to be $\kappa h_{\mu\nu}$, so that $h_{\mu\nu}$ is the canonically normalised field. We get the following linearised Lagrangian:
\be\label{lin-EH}
{\cal L}^{(2)}_{\rm EH}= -\frac{1}{2} (\partial_\sigma h_{\mu\nu})^2 + \frac{1}{2} (\partial_\mu h)^2 + (\partial^\nu h_{\mu\nu})^2 + h \partial^\mu\partial^\nu h_{\mu\nu},
\ee
where as usual we have introduced the notation $h:=g^{\mu\nu} h_{\mu\nu}$, with $g_{\mu\nu}$ being the background (Minkowski) metric. 

We note that (\ref{lin-EH}) is much more involved than the gauge-theory Lagrangians we have encountered above. We also note that the trace part $h$ of the metric perturbation behaves like a field with a wrong sign in front of its kinetic term. This is of course not a source of problems in GR, for this field turns out not to be dynamical. However, this does create problems for the path integral approach to the gravitational quantum theory, where one has to integrate over all metric fluctuations, including those described by $h$. This mode makes the Euclidean signature path integral not convergent. We will come back to this point below.

The above Lagrangian is invariant under the following transformations:
\be
\delta_\xi h_{\mu\nu} = \partial_{(\mu} \xi_{\nu)},
\ee
which represent linearised diffeomorphisms. 

The Hamiltonian description of the linearised GR is (briefly) as follows. Choosing the time+space split and decomposing the metric perturbation into its $h_{00}, h_{0i}, h_{ij}$ components we get (after quite some rearrangements and cancellations) the following involved structure:
\be\nonumber
{\cal L}^{(2)}_{\rm EH}=\frac{1}{2}(\dot{h}_{ij})^2 - \frac{1}{2}(\dot{h}_{kk})^2 - \frac{1}{2}(\partial_i h_{jk})^2+\frac{1}{2} (\partial_i h_{kk})^2 + (\partial^i h_{ij})^2 + h_{kk} \partial^i\partial^j h_{ij} \\ \nonumber + h_{00}(\Delta h_{kk}-\partial^i\partial^j h_{ij}) + 2h_{0i} \partial_j p^{ij},
\ee
where we have introduced:
\be
p^{ij}:=\dot{h}^{ij} - \delta^{ij} \dot{h}_{kk}.
\ee
It is clear that $p^{ij}$ is the momentum canonically conjugate to $h_{ij}$ and that the components $h_{00}, h_{0i}$ have vanishing conjugate momenta and are Lagrange multipliers for constraints. One can then see that it is possible to set the trace part of $p^{ij}$ by an action of a timelike diffeomorphism, and set the tracefree part $\tilde{h}_{ij}$ of $h_{ij}$ to be transeverse $\partial^i \tilde{h}_{ij}=0$ by an action of a spatial diffeomorhism. Imposing these gauge-fixing conditions we get the following simple Hamiltonian ${\cal H}^{(2)}_{\rm EH}=p^{ij} \dot{h}_{ij} - {\cal L}^{(2)}_{\rm EH}$:
\be\label{H-EH}
{\cal H}^{(2)}_{\rm EH} = \frac{1}{2} \left( (p^{ij})^2 + (\partial_i h_{jk})^2\right),
\ee
where now both the metric perturbation and the conjugate momentum are symmetric traceless transverse tensors. We do not give all the details of this involved analysis, them main point of this discussion being to illustrate how much more complicated the case of metric-based GR is as compared to our treatment in the previous subsection. We have also verified that exactly the same Hamiltonian appears from the usual metric-based analysis and from our linearised Lagrangian (\ref{L-GR}). The Hamiltonian (\ref{H-EH}) resulting from the metric-based treatment, as well as the fact that the dynamical fields are symmetric traceless transverse tensors, is the reason why the graviton is referred to as a spin two particle. Having derived the same description of the graviton from the gauge theory Lagrangian (\ref{L-GR}) we have thus shown that the gauge-theoretic description of a spin two particle is possible.

Having reviewed the usual linearised GR situation, we are ready to discuss the question of the relation between our gauge-theoretic and the metric-based approaches. We have seen that once all the constraints are imposed and all the gauge freedom is fixed, one gets exactly the same Hamiltonian in both cases, the one describing two propagating polarisations of the graviton. Using the technical jargon, we can say that on-shell the two descriptions of gravity are the same. The question is if they are the same off-shell, i.e. if there exists a field redefinition (possibly complicated) that maps (\ref{L-GR}) into (\ref{lin-EH}). We will now give arguments that prove that there is no such a map, and so the two descriptions are different off-shell. 

First, let us do a simple count of the number of components in the fields involved. The metric perturbation $h_{\mu\nu}$ contains 10 components, while the connection $a_\mu^i$ contains 12. However, in fact as we have seen the Lagrangian (\ref{L-GR}) depends only on 8 of the 12 components of $a_\mu^i$ as it is invariant under the shifts (\ref{diffeo}) in the field space. It is certainly not possible to express 10 components of the metric perturbation in terms of 8 components on which (\ref{L-GR}) depends.

Another reason why the Lagrangians (\ref{L-GR}) and (\ref{lin-EH}) cannot be related is that they have very different properties in the field space. It is easiest to see this in the case when one passes to the Euclidean signature. Then there is no factors of $\im$ in the gauge-theoretic description, and the full connection $a_\mu^i$ is real. The self-dual two-forms in this case are also real. Then the gauge-theoretic Lagrangian is just the trace of the square of the real symmetric tracefree matrix $P^{ij|kl}\Sigma^{\mu\nu}_k \partial_\mu a_{\nu l}$. As such, it is explicitly non-negative. On the other hand, the Lagrangian (\ref{lin-EH}), even when continued to the Euclidean signature, is still not positive-definite. Indeed, by choosing the conformal mode part $h$ of the metric perturbation to be a field quickly changing in spacetime one can make the Euclidean Lagrangian arbitrarily negative. This is well-known as the conformal mode problem in the Euclidean path integral. Thus, we are dealing with very different functionals in the two descriptions. In the gauge-theoretic description the (Euclidean) Lagrangian is convex (apart from the flat directions corresponding to the gauge symmetries), while in the metric-based treatment the functional is convex in some directions in the field space and concave in others (specifically the conformal mode direction). The critical point(s) of both functionals are the same, but it is clear that there cannot be any map from the field space of one description to that of the other that can relate two functionals with completely different properties. Thus, we conclude that the above gauge-theoretic description of linearised gravity is off-shell different from the description given by the Einstein-Hilbert action. This fact will have important implications when we discuss the quantum theory. 

\subsection{Spinor description}

The above description of the linearized theory, as well as the comparison to the metric-based approach, can be simplified quite considerably if one uses spinors. The main idea of the spinor description is that the space $S_+\otimes S_-$, where $S_\pm$ are the spaces of the so-called unprimed and primed spinors (inequivalent two-dimensional representations of the Lorentz group), is isomorphic to the Minkowski space. This isomorphism is described by the object $\theta_\mu^{AA'}$, where we use the relativist notations for spinors and denote the spinor indices by upper case Latin letters from the beginning of the alphabet. The spinors $\lambda^{A'}$ of opposite type to $\lambda^A$ are called primed. This is in contrast to the particle physics notations where the spinor indices are Greek letters, and are referred to as undotted and dotted. The (hermitian in our conventions) object $\theta_\mu^{AA'}$ is called the soldering form. It encodes, in particular, the Minkowski metric via $g_{\mu\nu}=-\theta^{AA'}_\mu \theta_{\nu AA'}$, where the minus in this formula is convention (and signature) dependent. Using the soldering forms one can convert any spacetime index into a pair of spinor indices of opposite types. Thus, for example the gauge field $a_\mu^a$ becomes $a_{AA'}^a$, where $a$ is the Lie algebra index. For more information about spinors and their role in geometry the reader is referred to \cite{Penrose:1985jw}.

We now describe the linearized Lagrangian (\ref{L-GR}) in spinor terms. As we shall see, most of the properties discussed above become manifest in this description. We already know that the spacetime index of the connection $a_\mu^i$ becomes a pair $AA'$. We then recall that the self-dual two-forms $\Sigma^i_{\mu\nu}$ provide an isomorphism between the space of self-dual two-forms and the Lie algebra of ${\rm SO}(3)$, and take in the spinor language a very simple form. Indeed, a two-form $X_{\mu\nu}$ becomes in the spinor description an object $X_{AA' BB'}$. This is anti-symmetric under the exchange of pairs $AA'$ and $BB'$. As such, it can be decomposed into either an object that is symmetric in $AB$ and anti-symmetric in $A'B'$, or the object anti-symmetric in $AB$ and symmetric in $A'B'$. Using the fact that the only $AB$ anti-symmetric object is the metric in the space of spinors $\epsilon_{AB}$ we have:
\be
X_{AA' BB'}= \epsilon_{A'B'} X_{AB} + \epsilon_{AB} X_{A'B'},
\ee
where $X_{AB}$ and $X_{A'B'}$ are symmetric objects. Then it can be shown that the $\epsilon_{A'B'} X_{AB}$ part of $X_{\mu\nu}$ is its self-dual part, and $\epsilon_{AB}X_{A'B'}$ is the anti-self-dual one. The objects $\Sigma^i_{\mu\nu}$ being self-dual are thus of the form $\epsilon_{A'B'} \Sigma^i_{AB}$. We can then use the objects $\Sigma_{AB}^i$ to identify the Lie algebra index $i$ with a symmetric pair $AB$ of unprimed spinor indices. All in all, when all of the indices are replaced by the spinor ones, our linearized connection $a_\mu^i$ becomes the object $a^{BC}{}_{AA'}$. 

Thus, we see that the linearized connection $a^{BC}{}_{AA'}$ is an object taking values in $S_+^2\otimes S_+\otimes S_-$, where the notation is that $S_+^n$ is the space of symmetric rank $n$ unprimed spinors. This space can be decomposed into its irreducible components:
\be
S_+^2\otimes S_+\otimes S_- = S_+^3 \otimes S_- \oplus S_+\otimes S_-,
\ee
where the completely symmetric $a^{(ABC)}{}_{A'}$ in its three unprimed indices part of $a^{BC}{}_{AA'}$ lies in $S_+^3\otimes S_-$ and the trace part $a^{EA}{}_{EA'}$ lies in $S_+\otimes S_-$. It is then easy to see that the part of the connection that can be shifted by the action of a diffeomorphism $\delta_\xi a_\mu^i=\xi^\alpha \Sigma^i_{\mu\alpha}$ is precisely the $S_+\otimes S_-$ part. And indeed, the linearized Lagrangian (\ref{L-GR}) written in terms of spinors take the following extremely simple form:
\be\label{L-GR-spinor}
{\cal L}^{(2)}_{\rm GR}\sim (\partial^{(A}_{A'} a^{BCD)A'})^2,
\ee
where the numerical coefficient in front is convention dependent and is not of importance for us here. Since the symmetrization on all 4 unprimed indices is taken, the above Lagrangian is explicitly independent of the $S_+\otimes S_-$ part of the connection. It is also easy to check that the Lagrangian (\ref{L-GR-spinor}) is invariant under the gauge transformations $\delta_\phi a^{BC}{}_{AA'}=\partial_{AA'} \phi^{BC}$. Indeed, this follows from the fact that $\partial^{AA'}\partial^B_{A'}\sim \epsilon^{AB} \Box$ and thus contains only the $AB$ anti-symmetric part, which gets killed by the symmetrization in (\ref{L-GR-spinor}). 

It is worth emphasizing how much simpler is the description (\ref{L-GR-spinor}) as compared to all other known descriptions of linearized gravity. We shall give a comparison to the Plebanski-Ashtekar theory in the following subsection. Here we would like to present a brief comparison to the spinor language metric case. In the spinor description the metric perturbation $h_{\mu\nu}$ becomes $h_{AA' BB'}$, which is symmetric under the exchange of pairs $AA'$ and $BB'$. As such, it decomposes into an object $h_{AB A'B'}\in S_+^2\otimes S_-^2$, as well as the object $\epsilon_{AB}\epsilon_{A'B'} h$ proportional to the trace of $h_{\mu\nu}$. Thus, there is $9+1=10$ field components, and the Lagrangian (\ref{lin-EH}) is a (complicated) function of all these fields. We note that even writing the Lagrangian (\ref{lin-EH}) in the spinor form produces a rather cumbersome expression that is not particularly illuminating. 

We have already remarked that the gauge-theory gravitational Lagrangian (\ref{L-GR}) is very similar to the Yang-Mills Lagrangian in the form (\ref{L-YM-S}). This similarity becomes even more striking in the spinor description. Thus, suppressing the Lie algebra index of the Yang-Mills field we can write (\ref{L-YM-S}) as
\be\label{L-YM-spinor}
{\cal L}^{(2)}_{\rm YM}\sim (\partial^{(A}_{A'} a^{B)A'})^2.
\ee
Thus, the only difference between this Yang-Mills theory and (\ref{L-GR-spinor}) is that the former uses spin one fields in the representation $S_+\otimes S_-$ (per Lie algebra generator), while the latter uses spin two fields in the representation $S_+^3\otimes S_-$. Otherwise, both Lagrangians are constructed according to the same principle using the Dirac operator $\partial_{AA'}$. Indeed, in (\ref{L-YM-spinor}) the Dirac operator is used to map $\partial: S_+\otimes S_-\to S_+^2$, and this symmetric rank two spinor is just squared to obtain the Lagrangian (and then the trace over the underlying Lie algebra is taken). In the case of (\ref{L-GR-spinor}) the principle is exactly the same except that the map is now $\partial: S_+^3\otimes S_-\to S_+^4$, as is relevant for spin two particles. Below we shall see that the similarity between this description of gravity and Yang-Mills theory extends beyond the linearized level, and that interaction vertices in our gauge theory description of gravity take very similar form to those familiar from the Yang-Mills. 

Let us now see how the count of the number of degrees of freedom works in the metric and the gauge descriptions. The first difference is that the gauge theory Lagrangian is a function of only 8 fields taking values in $S_+^3\otimes S_-$, as compared to 10 fields from $S_+^2\otimes S_-^2\oplus ({\rm trivial})$ in the metric case. In this sense, the gauge theory description is more economical than the metric-based one. The count of the propagating modes is then as follows. From the 8 fields that the Lagrangian (\ref{L-GR-spinor}) depends on, 3 becomes Lagrange multipliers for 3 Gauss constraints, which gives the usual 2 propagating degrees of freedom. This is to be contrasted with the metric case count, where 4 out of 10 metric components are Lagrange multipliers for 4 constraints, which again give two propagating modes. The point that we would like to emphasize is that the gauge-theory description is less redundant -- a smaller number of fields than in the metric case is sufficient to produce the same on-shell picture. 

Second, the Lagrangian (\ref{L-GR-spinor}) is arguably a very simple and natural construct involving the basic field, in comparison with the metric Lagrangian that is not even easy to write down with all its 4 different terms. The final, crucial difference comes from the fact that the Euclidean version of the Lagrangian (\ref{L-GR-spinor}) is positive definite (being the trace of the square of a real symmetric tracefree matrix), while the Euclidean version of the linearized Einstein-Hilbert functional is not definite (because of the "conformal" mode $h$). As we have already discussed above, this shows that the two descriptions can only be equivalent on-shell, while are necessarily different off-shell. 

Thus, all in all, the gauge theory formulation (\ref{L-GR-spinor}) seems to be a much more economical and elegant description of gravity than the usual metric based one. There is, however, a price to be paid for this simplicity. Indeed, the fields of the metric-based description take values in $S_+^2\otimes S_-^2$ plus the trivial representation of the Lorentz group. In Lorentzian signature, the representations $S_+$ and $S_-$ go into each other under the complex conjugation. Thus, the representation $S_+^2\otimes S_-^2$ goes into itself under the complex conjugation, which explains why the metric-based description works with real fields. In contrast, in our approach the basic field take values in $S_+^3\otimes S_-$, which under the operation of complex conjugation goes to $S_-^3\otimes S_+$, a completely different representation. In other words, our description of gravity is necessarily chiral, with gravitons of one helicity being described in a different way from those of the other helicity. The chirality also implies that the gauge theory description cannot be by real fields (apart from the cases of Euclidean $(+,+,+,+)$ and split $(-,-,+,+)$ signatures where the spinors are real objects). As the result, the issue of reality conditions becomes non-trivial. 

\subsection{A comparison to the Plebanski-Ashtekar gauge-theoretic description}

We close this section with a comparison between the above gauge theory description and that due to Plebanski \cite{Plebanski:1977zz} and (in the Hamiltonian formulation) Ashtekar \cite{Ashtekar:1987gu}. A detailed description of the corresponding linearized theory is available in e.g. \cite{Ashtekar:1991hf}, Chapter 11. Here we will only give a very brief comparison, for more details the reader is referred to the above references.

In Plebanski-Ashtekar description one starts from a certain action functional that depends on an ${\rm SO}(3)$ connection $A_\mu^i$, as well as an ${\mathfrak su}(2)$ Lie algebra valued two-forms field $B^i_{\mu\nu}$, as well as a set of Lagrange multipliers. The precise form of this functional is not important for us here. What matters is the fact that it is not a functional of the connection only, as in our description, in that it depends on many other fields. In the Hamiltonian description many of these fields become non-dynamical. Thus, as usual, the temporal part of the connection becomes the Lagrange multiplier for the Gauss constraint. The temporal part $B^i_{0j}$ of the B-field contains 9 components, of which 5 are set to zero by constraints obtained by varying the action with respect to the Lagrange multipliers, and the remaining 4 are the Lagrange multipliers for the diffeomorphism constraints. The spatial part $b^i_{jk}\sim \epsilon_{jkl} \pi^{li}$ of the two-form field plays the role of the momentum conjugate to the spatial connection. The linearized theory constraints are then as follows:
\be\nonumber
\partial_i \pi^{ij} = 0 \quad ({\rm Gauss}), \\  \label{Ashtekar-constr}
\partial_{[i} a_{j]}^i = 0 \quad ({\rm Diffeo}), \\ \nonumber
\epsilon^{ijk} \partial_i a_{jk} = 0 \quad ({\rm Hamiltonian}),
\ee
where $a_i^j$ is the spatial connection. The analysis of the propagating mode content of this system is quite non-trivial (it can be found in \cite{Ashtekar:1991hf}, Chapter 11), with the outcome being that the usual symmetric traceless transverse tensors are the propagating degrees of freedom. 

Let us now compare the above to our gauge theoretic description. The first difference is that our Lagrangian is a functional of the connection only, with no other auxiliary fields present. The second (crucial) difference is that the diffeomorphisms are realized in our description in a completely different way. Indeed, the spatial diffeomorphism as well as the Hamiltonian constraints (\ref{Ashtekar-constr}) are non-trivial, and follow from varying the Plebanski action with respect to certain components of the temporal part of the B-field. In contrast, in our description there are no fields in the action variation with respect to which produces the diffeomorphism constraints. Instead, the action is simply independent of certain components of the connection. At the level of the Hamiltonian description this is described by the statement that the following constraints hold: 
\be\label{diffeo-constr}
\pi^{[ij]}=0 \quad ({\rm Diffeo}), \quad \pi^{ii}=0 \quad ({\rm Hamiltonian}).
\ee
As we have already said, these follow from impossibility to solve for certain components of the velocity in terms of the momenta, and not by explicitly varying the Lagrangian with respect to some Lagrange multipliers. What the constraints (\ref{diffeo-constr}) generate are of course simple shifts of the connection variable in some directions. As the result, the analysis of the propagating mode content of (\ref{L-GR}) is much simpler than the corresponding analysis in the Ashtekar case. 

All in all, our gauge theory description is considerably different from the Plebanski-Ashtekar formulation, with both the linearized Lagrangian and Hamiltonian formulations being much more compact in our case. It is also worth emphasizing that the Plebanski Lagrangian, being essentially the Einstein-Hilbert Lagrangian with some extra fields added, is non-convex, while the (Euclidean) linearized Lagrangian (\ref{L-GR}) is explicitly convex. Thus, there are clear benefits from using the description (\ref{L-GR}) as compared to the Plebanski-Ashtekar formulation. 

Finally, we note that our Lagrangian (\ref{L-GR}) can be obtained from that of Plebanski theory by the process of integrating out the two-form and the Lagrange multipliers fields, as was explained in \cite{Krasnov:2011pp}, and then considering the linearization of the resulting action, as will be explained below. So, Plebanski formulation of GR and the formulation that is the subject of this review are not unrelated. However, we will refrain from explaining this link in details, as this will take us too far from the main subject of this paper, which is how to describe gravity by a theory with a connection as the only dynamical field.

\section{Diffeomorphism invariant gauge theories}
\label{sec:digt}

In this section we finally explain where the postulated above linearized gravity Lagrangian (\ref{L-GR}) comes from. We will also explain how both gravity and Yang-Mills theory arise from a single source - a general diffeomorphism invariant gauge theory. In this section we switch gears completely, and follow a top-to-bottom approach, where we first present a general principle, and then specialise to the case of gravity plus Yang-Mills. 

Some historical remarks are in order. While the gauge theories we encounter in this section are relatively new -- in the form described here they have been introduced in works of the present author -- they are intimately related to certain generally-covariant theories of connection that has been discovered and studied in the early 90's. In particular, the constructions of this section will make it clear that GR is not the only interacting theory of massless spin two particles. This fact has been known for quite some time in a different, but not unrelated formulation. The history of these developments is (briefly) as follows. It was realised in work \cite{Capovilla:1989ac} (see also \cite{Capovilla:1991kx}) that the zero cosmological constant GR can be reformulated as a theory of an ${\rm SU}(2)$ connection. The Lagrangian of this description, however, contains not just the connection field, but also an additional auxiliary field of density minus one. It was then quickly realised that in this formulation GR is not unique. Thus, a two parameter family of "neighbors" of GR was introduced and studied in \cite{Capovilla:1992ep}, and then an infinite parameter of theories constructed along the same lines was proposed in \cite{Bengtsson:1990qg}. All these theories are not "pure connection" ones since, as we have already said, they contain an additional auxiliary field. But they all have the key property that they describe just two propagating polarisations of the graviton. The same theories were rediscovered (in a different formulation) in \cite{Krasnov:2006du} and this new version was related to earlier developments in \cite{Bengtsson:2007zzd}. The novelty of the approach followed here is that Lagrangians that are used are functions of only the connection field, with no auxiliary field being necessary. However, our theories can be though of as those studied in the 90's with the auxiliary field integrated out. 

\subsection{A class of gauge theories}

Let us start by describing how gauge theory actions can be constructed in the absence of any spacetime metric. This question is non-trivial, for up to now we only know how to formulate a gauge theory (i.e. Yang-Mills theory) when a spacetime metric is available (actually, since Yang-Mills is classically conformally-invariant only a conformal class of the metric is needed). It is not at all clear how the spacetime indices of the curvature two-forms can be contracted in the absence of any metric. 

To begin to answer this question we note that the only object that can be used to contract the indices of copies of $F^a_{\mu\nu}$ and that is available without a metric is the tensor density $\tilde{\epsilon}^{\mu\nu\rho\sigma}$. This is a completely anti-symemtric object that in any coordinate system has components $\pm 1$. We thus see that the following object with all spacetime indices contracted is available:
\be\label{X}
\tilde{X}^{ab}:=\frac{1}{4} \tilde{\epsilon}^{\mu\nu\rho\sigma} F^a_{\mu\nu} F^b_{\rho\sigma},
\ee
where we have explicitly indicated that $\tilde{X}^{ab}$ is a density weight one (by putting a tilde over the symbol). The numerical prefactor of $1/4$ is introduced for convenience so that with our conventions $F^a\wedge F^b=\tilde{X}^{ab} d^4 x$. We now construct our Lagrangian by considering the most general gauge invariant quantity that can be produced from (\ref{X}). 

Thus, the task is now to construct a scalar from (\ref{X}). Moreover, this scalar must be a density one, so that it can be integrated over the spacetime to produce an action. Thus, what we require is a function from the space ${\mathfrak g}\otimes_S{\mathfrak g}$, where $\mathfrak g$ is the Lie algebra of $G$ and $S$ denotes the symmetrisation, to the real (or possibly complex) numbers, and this function must be homogeneous of degree one:
\be
f: {\mathfrak g}\otimes_S{\mathfrak g} \to \R (\C), \qquad f(\alpha X)=\alpha f(X) \quad \forall X\in {\mathfrak g}\otimes_S{\mathfrak g}, \forall \alpha \in \C^*,
\ee
where the last condition comes from our desire for $f(\tilde{X})$ to be a density weight one. We also want our action to be gauge-invariant, and so since the gauge transformations act on the curvature two-forms by rotating the Lie algebra indices $a$ (using mathematical terminology, the adjoint action), we also require
\be
f( {\rm ad}_g X) = f(X) \quad \forall X\in {\mathfrak g}\otimes_S{\mathfrak g}, \forall g\in G,
\ee
where ${\rm ad}_g X$ is the extension of the adjoint action of $G$ in $\mathfrak g$ to ${\mathfrak g}\otimes_S{\mathfrak g}$.

One can then easily convince oneself that the class of such functions is not empty by giving examples. The first obvious example is obtained by taking $f(X)={\rm Tr}(X)$. However, it is easy to see that the corresponding Lagrangian is a total derivative, and thus does not give rise to an interesting dynamical theory. A more non-trivial example is $f(X)={\rm Tr}(X^2)/{\rm Tr(X)}$, which is homogeneous of degree one as required and gauge-invariant. In general, one can construct a very large of functions with desired properties by taking all possible (independent) invariants constructed from $X\in {\mathfrak g}\otimes_S{\mathfrak g}$, and then taking an arbitrary functions of homogeneity degree zero ratios of such invariants, and multiplying it by, say, ${\rm Tr}(X)$ to get the required total homogeneity degree one. For our purposes here it is enough to know that a very large class of functions with required properties exists. 

Having discussed the properties of functions that can be used in constructing the action, we are ready to write down an action principle for a diffeomoprhism invariant gauge theory. Using the form notations we have
\be\label{action}
S[A]=\im\int f(F\wedge F),
\ee
where it is understood that the wedge product of two curvature two-forms is 4-form valued in ${\mathfrak g}\otimes_S{\mathfrak g}$, and the properties of function $f$ make $f(F\wedge F)$ a well-defined 4-form that can be integrated over the spacetime manifold to get the action. The imaginary unit $\im$ is introduced in front of the action for future convenience.

For future use, we note that the field equations following from (\ref{action}) when this action is varied with respect to the connection are:
\be\label{feqs}
d_A \left( \frac{\partial f}{\partial \tilde{X}^{ab}} F^b \right)=0.
\ee
Here $d_A$ is the covariant exterior derivative with respect to the connection $A$, and $\partial f/\partial \tilde{X}^{ab}$ is the matrix of partial derivatives of the function $f$ with respect to the components of its matrix argument $\tilde{X}^{ab}$. We note that the matrix of first partial derivatives is a function of $\tilde{X}$ that is homogeneous of degree zero. It is worth emphasizing that the field equations (\ref{feqs}) are (non-linear) partial differential equations that are second order in derivatives. The fact that only second order differential equations appear is reassuring, for higher order field equations typically lead to instabilities. 

It is instructive to compare (\ref{feqs}) to the field equations following from the Yang-Mills action functional. These read $d_A {}^* F^a=0$, where the dependence on the spacetime metric enters via the Hodge star operation applied to $F$. Thus, the equations (\ref{feqs}) are similar to those of the Yang-Mills theory, except that instead of applying to the curvature the Hodge star operator (which does not exist in our case), one contracts it with a matrix of first derivatives of the function $f$, which is itself depending on the curvature. Thus, equations (\ref{feqs}) are well-defined in the absence of any spacetime metric.  

Importantly, we note that there are no dimensionful constants involved in the definition of the theory (\ref{action}). Indeed, it is natural to take the dimensions of the connections to be those of $1/L$, $L$ being length (or, using the standard terminology mass dimension one). The quantities $\tilde{X}^{ab}$ are then of mass dimension 4, and due to the homogneity of $f$, so is the Lagrangian. Thus, there are only dimensionless constants involved in the construction of the Lagrangian of our theory, and these are hidden as the parameters of the function $f$.

Our final remark is about the generality of the above construction of a diffeomorphism invariant gauge theory. We have considered actions that are built from the quantities (\ref{X}). The natural question to ask is if there are any more involved gauge and diffeomorphism invariant functionals of the connection that can be constructed. As we shall discuss in more details below, this question is very important for understanding of the quantum theory based on (\ref{action}), for it is related to the question of how the class of theories (\ref{action}) behaves under the renormalization. To discuss this question, we note that a certain "metric" tensor can be produced out of the curvature two-forms via:
\be\label{Urbantke}
\tilde{g}_{\mu\nu} = \tilde{\epsilon}^{\alpha\beta\gamma\delta} g_{ab} C^a_{cd} F^c_{\mu\alpha} F^d_{\nu\beta} F^b_{\gamma\delta}.
\ee
Here $C^a_{cd}$ are the structure constants, and $g_{ab}$ is the Killing-Cartan metric on the Lie algebra $\mathfrak g$. The object $\tilde{g}_{\mu\nu}$ is a tensor density of weight one, symmetric under the exchange of $\mu\nu$. It is a general group $G$ analog of the so-called Urbantke metric \cite{Urbantke:1984eb}. The choice of the proportionality coefficient in this formula is rather arbitrary, and one can e.g. always divide the object (\ref{Urbantke}) by ${\rm Tr}(\tilde{X})$ to get an object of density weight zero. So, in general (\ref{Urbantke}) is only defined modulo rescalings. However, what is important for us here is that this object exists. In general, it is a non-degenerate tensor, which thus has an inverse. This inverse can then be used to contract indices of the operator of the covariant derivative $d_{A\, \mu}$ with that of the curvature $F^a_{\rho\sigma}$, producing a Lie algebra valued object with one spacetime index. This object can in turn be used to construct gauge and diffeomorphism invariant objects of the type different from those considered before. Thus, the class of theories (\ref{action}) describes only a subset of diffeomorphism invariant gauge theories. This subset can be characterised by noting that for any actions other than (\ref{action}) the resulting field equations are of higher than second order in derivatives. Thus, the class (\ref{action}) consists of those diffeomorphism invariant theories that lead to not higher than second order field equations. It is therefore certainly justified to restrict one's attention to such theories at low energies. A discussion of issues arising at high energies, when also the quantum effects become important, will be given below. 

We would like to close this subsection with two references on somewhat analogous constructions of action functional that have appeared in the literature. Thus, reference \cite{Hitchin:2001rw} presents an action that is a functional of a $p$-form in $n$-dimensions. Such an action is constructed as a homogeneous of degree $n/p$ function of a form. The result is a well-defined $n$-form that can be integrated to produce a "volume" functional. The difference with our case is that we have constructed an action from Lie-algebra valued 2-forms (in 4 spacetime dimensions), while in \cite{Hitchin:2001rw} non-trivial functionals arise for higher-rank forms (3- and 4-forms) in higher dimensions. The forms in \cite{Hitchin:2001rw} are usual differential forms, not forms with values in vector bundles as in our context.

Another relevant construction is \cite{Endlich:2010hf}, where an action principle for a perfect fluid is given in a form quite analogous to our (\ref{action}). Here one also works with symmetric matrices valued in some internal space, and constructs a gauge-invariant function of such matrices which is then integrated to produce an action. In addition, the requirement of invariance under the volume-preserving diffeomorphisms is imposed. The main difference with our context is that our basic field is a connection one-form, while in \cite{Endlich:2010hf} it is a spacetime scalar with values in some internal space. The perfect fluid construction of  \cite{Endlich:2010hf} suggests that the elementary excitations of our theory -- gravitons -- can be interpreted as analogs of phonons. Indeed, our gravitons arise as gapless modes once the original symmetries of the theory -- diffeomorphisms and gauge rotations -- are broken to a smaller (Poincare) subgroup by the background. We now turn to describing the linearization of (\ref{action}) around an appropriately chosen background. 

\subsection{Background}

We start by noting that the construction (\ref{action}) of a diffeomorphism invariant gauge theory is empty in the simplest case $G={\rm U}(1)$. Indeed, with the Lie algebra being one-dimensional, there is no other independent invariant of $\tilde{X}^{ab}$ except the trace (which in this case coincides with the quantity itself), and so the only possible theory in that case is the topological one (with the action being a total divergence). The next to most simple case is $G={\rm SO}(3)$, which we shall now see describes gravity. 

We will first explain how the Lagrangian (\ref{L-GR}) is obtained from (\ref{action}) when the theory is linearised around a specific background. We will see that for this calculation the specific form of the function $f$ defining the theory is not important. Any choice of $f$ satisfying a certain weak non-degeneracy condition will produce (\ref{L-GR}) via linearisation. Then in the next subsection we describe how a specific choice of $f$ gives GR. 

We now need to discuss the background connection to be used for the linearisation. We present a general description valid for any gauge group $G$, and then later specialize to the case $G={\rm SO}(3)$. We first note that because ratios of invariants of $\tilde{X}^{ab}$ are used, we cannot take the zero (flat) connection with vanishing curvature. Indeed, the ratios of curvature invariants that are arguments of the function $f$ are then ill-defined. This is probably the most significant difference with the case of Yang-Mills theory, where the theory is completely well-defined around the zero connection. In our case, to obtain an expansion around the Minkowski background, there is no other choice but to resort to some sort of limiting procedure. A natural limiting procedure is to take a constant curvature connection and then send the radius of curvature to infinity, thus reproducing the Minkowski spacetime. We shall see that this is a well-defined procedure with an unambiguous outcome. Alternatively, one can interpret this limiting procedure as that of finding the linearised description of the theory (\ref{action}) for a constant curvature background, on scales much smaller than the curvature scale. We note that this is a physically realistic situation, because the metric of our Universe is not Minkowski, and, during the current late stages of the evolution of the Universe, is in fact quite close to the de Sitter space.

To describe the background connection, we shall require it to have a large degree of symmetry, so that it can legitimately be referred to as a "vacuum" state of our theory. The symmetry requirements that we impose are those standard in cosmology: homogeneity and isotropy. Thus, we require that there exists a foliation of the spacetime such that on each 3-dimensional slice of the foliation the connection is homogeneous, i.e. looks exactly the same at every point. This is formalized by introducing a time coordinate, which we for now shall refer to as $\eta$, so that $\eta=const$ are the time-slices of homogeneity. The connection will then only be allowed to depend on $\eta$. We then require the connection to be spherically-symmetric, i.e. such that the effect of spatial rotation around any point on $\eta=const$ hypersurface can be removed by a gauge transformation. It is not hard to show that this fixes the connection to be of the form
\be\label{A-backgr}
A^a_{\rm backgr} = \frac{a^a_i}{\im} dx^i,
\ee
where $x^i$ are the usual $\R^3$ coordinates on $\eta=const$ surfaces, $a^a_i$ is a function of $\eta$ only, and we have used the availability of time-dependent gauge transformations to eliminate the $d\eta$ components. The requirement of spherical symmetry further requires $a^a_i$ to be an {\it embedding} of the Lie algebra ${\mathfrak so}(3)\sim {\mathfrak sl}(2)$ of the group of spatial rotations ${\rm SO}(3)$ into the Lie algebra $\mathfrak g$, i.e. a map that sends ${\mathfrak sl}(2)$ commutators into $\mathfrak g$ commutators. We refer the reader to \cite{Krasnov:2011hi} for a demonstration of this. In the case $G={\rm SO}(3)$ there is only one such (non-trivial) embedding, and so in this case $a^i_j$ is proportional to $\delta^i_j$, but in the case of a larger gauge group $G$ there are typically several inequivalent (i.e. non-conjugate in $G$) embeddings. Later we shall see that it is very interesting to consider different embeddings in (\ref{A-backgr}), as one gets very different particle spectra around non-equivalent "vacua". 

Let us now rewrite the connection components $a^a_i$ as 
\be
a^a_i = e^a_i a(\eta),
\ee
where $e^a_i$ is an embedding of ${\mathfrak sl}(2)$ into $\mathfrak g$ normalized so that $f^a{}_{bc} e^b_j e^c_k = \epsilon^i{}_{jk} e^a_i$, and $a(\eta)$ is an arbitrary function of time $\eta$. We shall now see that the only freedom in choosing the vacuum (\ref{A-backgr}) is the choice of the embedding, and that the time coordinate can always be conveniently chosen so that the dependence of $a$ on time is fixed. Thus, let us compute the curvature of (\ref{A-backgr}). We have
\be\label{F-backgr-1}
F^a = - a^2 e^a_i \left( \im \frac{a'}{a^2} d\eta \wedge dx^i + \frac{1}{2} \epsilon^i{}_{jk} dx^j \wedge dx^k\right). 
\ee
We now choose the time coordinate so that $a'/a^2 d\eta = dt$, which fixes the connection component $a$ as a function of $t$:
\be
a(t)=\frac{1}{t_0-t}, \qquad t<t_0,
\ee
where $t_0$ is an arbitrary integration constant. 

Let us now inspect the background curvature (\ref{F-backgr-1}) more closely. The expression (\ref{F-backgr-1}) shows, in particular, that in the 6-dimensional space of all two-forms the background curvature components span a three-dimensional subspace 
\be\label{Lambda-plus}
\Lambda^+ := {\rm Span}(F^a), \qquad {\rm dim}(\Lambda^+)=3.
\ee
We can now introduce a metric, or rather a conformal class of metrics, by requiring this subspace $\Lambda^+\subset \Lambda^2$ to be that of self-dual two-forms with respect to the metric. It is known that this requirement fixes the metric modulo conformal rescalings. Let us stress that we are not forced to introduce a metric, as our theories are perfectly well-defined and can be studied in the absence of one. However, since all the physics as we know it happens in the metric background, it is very convenient to introduce a metric so that later we have a chance to see the familiar physics arising. We are not yet claiming that the metric introduced via the above mathematical construction is a physical one, in the sense that it is the one that the matter fields feel. However, we shall later see that this is indeed the case. 

Let us now discuss the ambiguity in the choice of the representative in the conformal class defined by declaring $\Lambda^+$ to be the self-dual two-forms. Given any such metric, let $\theta^I$ be the corresponding tetrads: $ds^2 =\theta^I\otimes \theta^J \eta_{IJ}$, where $I,J=1,\ldots, 4$ and $\eta_{IJ}$ is the Minkowski metric. We can then define the "canonical" self-dual two-forms for this metric via
\be\label{Sigmas}
\Sigma^i := \im \theta^0\wedge \theta^i +\frac{1}{2} \epsilon^i{}_{jk} \theta^j\wedge \theta^k.
\ee
For metrics related by a conformal rescaling $g_{\mu\nu} \to \Omega^2 g_{\mu\nu}$, the corresponding self-dual two-forms $\Sigma^i$ are related via $\Sigma^i_{\mu\nu} \to \Omega^2 \Sigma_{\mu\nu}$. Thus, any metric whose self-dual two-forms $\Sigma$ are a multiple of
\be
\Sigma^i_{\rm Mink} := \im dt \wedge dx^i + \frac{1}{2} \epsilon^i{}_{jk} dx^j \wedge dx^k
\ee
is a possible metric in the conformal class fixed by (\ref{Lambda-plus}). It is, however, very convenient to work with a metric such that 
\be\label{F-backgr}
F^a = - M^2 e^a_i \Sigma^i,
\ee
where $M^2$ is a constant (i.e. time independent) parameter. This is the de Sitter metric 
\be\label{de-Sitter}
ds^2 = c^2(t)\left( - dt^2 + \sum_i (dx^i)^2\right),
\ee
of constant curvature $M^2$ or of the cosmological constant 
\be\label{M-Lambda}
M^2=\Lambda/3.
\ee
Here
\be
c(t)=\frac{1}{M(t_0-t)}.
\ee
One could also consider anti de Sitter space if one so wished by taking the parameter $M$ to be purely imaginary. However, as the observational evidence points to the existence of a positive cosmological constant we choose our background metric to be that of a de Sitter space. 

The introduced parameter $M$ of dimensions of mass is completely arbitrary, for one can always rescale the metric by a constant, while at the same time rescaling $M^2$ by the inverse constant without changing the background curvature $F^a$. We now remind the reader that our theories do not have any dimensionful constants. We will later see that once (\ref{action}) is expanded around backgrounds (\ref{A-backgr}) interpreted as equipped with a metric of constant curvature $M^2$, the dimensonful parameter $M$ serves as a seed for all other dimensionful parameters in the theory, such as masses (for some massive fields that will be discussed below), and such as dimensionful coupling constants (such as e.g. the graviton interaction strength). Thus, the parameter $M$ is in fact just a unit of mass for our theories. As such, it is inconsistent to ask about its value, for there is nothing else that this value can be compared with. 

All our backgrounds (\ref{A-backgr}) are constant curvature, in e.g. the sense of equation (\ref{F-backgr}). Thus, the most symmetric solution of our theory (\ref{action}) is naturally a Universe with a cosmological constant (in the sense of the de Sitter metric (\ref{de-Sitter})). In this sense, a non-zero cosmological constant does not need to be explained in our formulation of gravity, it comes as a given. Moreover, as we have already stressed above, it is logically inconsistent to ask about its value, for being the only dimensionful parameter there is nothing that this value can be compared to. The cosmological constant is just a unit in which all other quantities get expressed. 

The above discussion has very interesting implications. In our theories the cosmological constant $\Lambda$, together with the usual $c,\hbar$, becomes the fundamental constant, such that any other constant of Nature can be expressed as a multiple of some combination of $\Lambda,c,\hbar$. Thus, for our class of theories the cosmological constant $\Lambda$ plays the same role as the Newton constant plays conventionally. In fact, we shall see that the Newton constant that measures the strength of interaction of gravitons does get expressed as a multiple (necessarily very large) of $M^{-2}$. When put in this way, the famous cosmological constant problem of why $\Lambda$ is so small becomes in our case the question of why $M_p^2\sim 1/G$ is so large (as compared to $\Lambda$). We will come back to this question below. 

\subsection{Action evaluated on the background}

Before we proceed with our analysis of the theories (\ref{action}) linearized around the background (\ref{A-backgr}) we make a small detour and compute the value that the action takes when evaluated on (\ref{A-backgr}). This small computation is related to the discussed above issue of other physical parameters being expressed in terms of the cosmological constant. 

We have $F^a\wedge F^b = M^4 e^a_i e^b_j \Sigma^i\wedge \Sigma^j = 2\im M^4 e^a_i e^b_j \delta^{ij} \sqrt{-g} \, d^4x$, where $\sqrt{-g}=c^4(t)$ is the square root of the determinant of the de Sitter metric. So, we have
\be
S[A_{\rm backgr}]= - 2M^4 f(m) \int\sqrt{-g}\, d^4x.
\ee
where we have introduced a notation $m^{ab}:=e^a_i e^b_j \delta^{ij}$ for the pullback of the metric on ${\mathfrak sl}(2)$ into $\mathfrak g$ by the embedding $e^a_i$. Comparing this with the value of the Einstein-Hilbert action evaluated on the constant curvature metric we see that we must identify
\be
2M^4 f(m) = \frac{\Lambda}{8\pi G}.
\ee
In other words, the product $M^4 f(m)$ must be identified with the energy density of the cosmological constant. Now, using the above identification (\ref{M-Lambda}) we see that we must have
\be\label{f-m}
f(m) \sim M_p^2/M^2.
\ee 
This is a very large number $M_p^2/M^2\sim 10^{120}$. Below we shall see similar large numbers appearing if we want to get the usual graviton interaction strength from our approach. Thus, we learn that the functions $f$ that are capable of reproducing the realistic physics are quite special in the sense that the above sort of scales appear. We will come to this point when we discuss what the cosmological constant problem becomes within our approach.  

\subsection{General linearization}

To see how (\ref{L-GR}) arises, let us first derive the general linearization formula, without specifying to the $G={\rm SO}(3)$ case. Thus, we will continue to use $a,b,\ldots$ for the Lie algebra indices. Let now $A_\mu^a$ to be (an arbitrary for now) background connection and $\delta A_\mu^a$ be a perturbation. The first variation of (\ref{action}) reads:
\be
\delta S= 2\im \int \frac{\partial f}{\partial \tilde{X}^{ab}} F^a \wedge D_A \delta A^b.
\ee
The second variation reads:
\be\label{sec-var}
\delta^2 S= \im \int 4 \frac{\partial^2 f}{\partial \tilde{X}^{ab}\partial \tilde{X}^{cd}} (F^a \wedge D_A \delta A^b) (F^c \wedge D_A \delta A^d) \\ \nonumber + 2 \frac{\partial f}{\partial \tilde{X}^{ab}} (D_A \delta A^a \wedge D_A \delta A^b + F^a C^{bcd} \delta A^c \wedge \delta A^d ).
\ee
Here $C^{abc}$ are the structure constants. One must substitute the background curvature (\ref{F-backgr}) (as well as the background connection) into this second order action, and study the spectrum of the arising excitations. 

\subsection{Case $G={\rm SO}(3)$ -- linearised theory}

We now note that in view of (\ref{metricity}) the background value of the matrix $\tilde{X}^{ij}$ is proportional to $\delta^{ij}$. Using this fact we can say a great deal about the matrices of first and second partial derivatives of $f$ present in (\ref{sec-var}). First, it is clear that the matrix of first derivatives can only be proportional to $\delta^{ij}$. To see this one can think about the function $f(\tilde{X})$ as being a power series expansion in traces of powers of $\tilde{X}$, as well as inverses of such traces. It is then clear that the value the matrix of first derivatives of such a function can take on $\delta^{ij}$ must itself be proportional to $\delta^{ij}$. The proportionality coefficient is then a constant (because the background was chosen to be of constant curvature). We can then integrate by parts in the first term on the second line in (\ref{sec-var}) and, assuming that the connection perturbation vanishes outside of the region of interest, see that the result cancels with the second term. So, for the background considered we can ignore the second line in (\ref{sec-var}). 

Let us now consider the term involving the matrix of second derivatives of $f$. The form of this matrix can be determined from the fact that $f$ is a homogeneous function of order one. This property can be summarised by writing
\be
\frac{\partial f}{\partial \tilde{X}^{ij}} \tilde{X}^{ij} = f.
\ee
This identity can be differentiated with respect to $\tilde{X}^{kl}$ with the result being
\be
\frac{\partial^2 f}{\partial \tilde{X}^{ij}\partial \tilde{X}^{kl}} \tilde{X}^{ij} = 0.
\ee
When the background value of $\tilde{X}^{ij}$ is proportional to $\delta^{ij}$ this implies that the matrix of second derivatives of $f$ gives zero when contracted with the Kronecker delta in any of its pairs of indices. This implies that the matrix of second derivatives is proportional to the projector on the symmetric tracefree tensors:
\be\label{f-2}
\frac{\partial^2 f}{\partial \tilde{X}^{ij}\partial \tilde{X}^{kl}} \Big|_{\tilde{X}=\delta} = -\frac{g}{2} P^{ij|kl}.
\ee
where $g$ is some constant, capturing the content of the Hessian of the function $f$ at $\tilde{X}^{ij}= \delta^{ij}$, and the numerical factor and the sign are introduced for convenience (we will later see that $g$ is positive for general relativity). Note that $g=0$ for the function $f(X)={\rm Tr}(X)$ corresponding to the topological theory, but for a generic choice of $f$ it is non-zero. The Hessian at the point $\tilde{X}^{ij} = 2\im M^4 \sqrt{-g} \, \delta^{ij}$ is easily determined using the homogeneity of $f$. All in all, collecting all the factors and dividing the second variation of the action by two to get the linearised action we get:
\be\label{S-2}
S^{(2)} = \frac{g}{2} \int \sqrt{-g} \, P^{ij|kl} (\Sigma_i^{\mu\nu} D_\mu \delta A_{\nu j}) (\Sigma_k^{\rho\sigma} D_\rho \delta A_{\sigma l}).
\ee
Note that factors of $M$ have cancelled. Now, to get the Lagrangian (\ref{L-GR}) we define a new, canonically normalised field $a_\mu^i = \im \sqrt{g} \delta A_\mu^i$. Note that there is a factor of the imaginary unit in this formula, as in e.g. (\ref{A-backgr}). We now take the limit $M\to 0$, or, alternatively, consider the above linearised theory on scales much smaller than the scale of the curvature. In both cases we can replace the covariant derivative present in (\ref{S-2}) by the ordinary derivative, and remove the volume density for the metric. We get:
\be\label{S-lin-gr}
S^{(2)} = -\frac{1}{2} \int  P^{ij|kl} (\Sigma_i^{\mu\nu} \partial_\mu a_{\nu j}) (\Sigma_k^{\rho\sigma} \partial_\rho a_{\sigma l}),
\ee
which is the integral of the Lagrangian (\ref{L-GR}) studied above. 

\subsection{Case $G={\rm SO}(3)$ -- recovering full GR}

We have just seen how for $G={\rm SO}(3)$ the linearisation of (\ref{action}) around the constant curvature background (\ref{A-backgr}) produces the gauge-theoretic gravity Lagrangian (\ref{L-GR}) studied in the previous section. This has happened independently of which function $f$ was taken in the construction of the action. We shall now show that for a particular choice of $f$ the full on-shell content of (\ref{action}) is the same as in GR. In other words, we will now show that for a specific $f$ there is a correspondence between solutions of general relativity and solutions of theory (\ref{action}). As we have already discussed in the section on the linearised theory, we can at best hope to have an on-shell correspondence between the two descriptions, for the off-shell even the linearised theories have completely different convexity properties. 

Let us first state that the function $f$ that corresponds to GR is given by \cite{Krasnov:2011pp}:
\be\label{f-GR}
f_{\rm GR}(X)= \frac{1}{16\pi G\Lambda} \left({\rm Tr}\sqrt{X}\right)^2,
\ee
where $G$ is the Newton's and $\Lambda$ is the cosmological constant. Note that, as discussed above, there are only dimensionless constants entering into the construction of $f$. We also note that the function $f$ evaluated on the identity matrix $X^{ij}=\delta^{ij}$ is of the order $M_p^2/M^2$, which we knew we should have expected from general considerations leading to (\ref{f-m}). We also note that the cosmological constant $\Lambda$ enters $f$ in the denominator, so the function $f$ as well as the action blow up in the limit $\Lambda\to 0$. Thus, strictly speaking, only the non-zero $\Lambda$ case GR is covered by our gauge-theoretic description. However, the flat case can be obtained by a limiting procedure of the sort we have used in the previous subsection\footnote{A "pure connection" action functional that covers the $\Lambda=0$ case is also possible, and is known as the Capovilla-Dell-Jacobson (CDJ) action \cite{Capovilla:1989ac}. It is, however, not of the type (\ref{action}) as it contains an additional auxiliary field. The CDJ action can be thought of as the $\delta$-function imposing the condition that the trace of the square root of the matrix of wedge products of the curvature vanishes. The same condition follows from (\ref{f-GR}) in the limit $\Lambda\to 0$, and in this sense the $\Lambda=0$ case is covered by our description via a limiting procedure.}. We feel that it is not a drawback of our scheme that only the non-zero $\Lambda$ case GR admits an honest action principle. Indeed, it is now commonly believed that there is a small cosmological constant in the Universe (or at least all the observational data are compatible with the dark energy being a cosmological constant). Thus, the action principle with $f$ as above is sufficient to describe the physical Universe. Also, as we have said, if desired, the properties of the flat Universe can be recovered by a limiting procedure. In addition, as we shall indicate below, the present gauge-theoretic formulation may allow to explain the form of the "physical" function $f$ as an outcome of some sort of renormalisation group flow. Thus, the present approach may have something to say about the problem why $\Lambda$ is so small --- the famous cosmological constant problem. More remarks on this will be given below. 

The notion of the square root used in (\ref{f-GR}) requires clarifications. As it can be shown, see \cite{Krasnov:2011pp} for more details, in the case of Euclidean signature general relativity the matrices $X$ that arise are squares of real matrices and therefore are real and not indefinite, i.e. all their eigenvalues are of the same sign (with some possibly being zero). For such matrices there is a well-defined notion of the square root, and this is what is used in (\ref{f-GR}). In the case of the physical, Lorentzian signature GR the situation is more complicated, because the matrices $X$ are in general complex. However, even in this case we can define the square root in the neighbourhood of the identity matrix by writing $\sqrt{X}=\sqrt{\delta + (X-\delta)}$ and expanding in powers of the small matrix $X-\delta$. This definition is sufficient for most practical applications, and certainly for the perturbative treatment of gravity. Thus, one can take the position that the function (\ref{f-GR}) can be defined by a power series expansion for curvatures whose departure from the constant curvature case is not large. And in the large curvature case where a perturbative definition would break down one cannot trust the theory of general relativity anyway, since higher order corrections to the GR Lagrangian (induced by e.g. quantum effects) must be taken into account. To summarise, the function (\ref{f-GR}) is well-defined in all situations where one is interested in classical GR predictions.

To show that the theory with (\ref{f-GR}) is on-shell equivalent to the usual GR with its Einstein-Hilbert action, let us first write down the field equations (\ref{feqs}), specialised to the case of the defining function $f$ given by (\ref{f-GR}). Let us first define, for any $f$
\be\label{B}
B^i :=\frac{\partial f}{\partial \tilde{X}^{ij}} F^j.
\ee
The field equations (\ref{feqs}) are then $d_A B^i=0$. For the function (\ref{f-GR}) we have
\be
\frac{\partial f}{\partial \tilde{X}^{ij}} = \frac{1}{16\pi G\Lambda} \left({\rm Tr}\sqrt{\tilde{X}}\right) \left(\tilde{X}^{-1/2}\right)^{ij},
\ee
where a negative power of the matrix $\tilde{X}^{ij}$ should also be understood in the sense of a power series expansion. It is then easy to check that precisely for the function (\ref{f-GR}) we have the following important identity satisfied by the two-forms $B^i$:
\be\label{B-metr}
B^i\wedge B^j \sim \delta^{ij}.
\ee
One can reverse this argument and say that the function $f$ in (\ref{f-GR}) is chosen precisely in such a way that the identity (\ref{B-metr}) holds. This identity is analogous to (\ref{metricity}) already encountered above. 

Overall, we get the following structure. First, any ${\rm SO}(3)$ connection $A^i$ gives via (\ref{B}) a set of two-forms $B^i$ satisfying (\ref{B-metr}). It is then known that such a triple of two-forms defines a unique spacetime metric $g_{\mu\nu}$. The metric $g_{\mu\nu}$ is determined by the requirement that the triple of two-forms $B^i$ is self-dual with respect to $g_{\mu\nu}$ (this defines the metric modulo conformal transformations), and that the volume form of the metric coincides with a multiple of ${\rm Tr}(B\wedge B)$. For an explicit expression, one can use the Urbantke formula (\ref{Urbantke}), specialized to the case ${\mathfrak g}={\mathfrak su}(2)$, with the two-forms $B^i$ substituted in place of $F^i$. Then for a connection satisfying its field equation $d_A B^i=0$ it can be shown that $A^i$ coincides with the self-dual part of the Levi-Civita connection for the metric defined by $B^i$. Finally, the defining relation (\ref{B}) can be read as saying that the curvature of the self-dual part of the Levi-Civita connection is self-dual as a two-forms, which is known to be equivalent to the Einstein condition, see e.g. Proposition 2.2. in \cite{Atiyah:1978wi}, or \cite{Krasnov:2009pu} for an exposition oriented towards a physics audience. This establishes the on-shell equivalence of theory (\ref{action}) with (\ref{f-GR}) with Einstein's general relativity. As we have already discussed, there is only the on-shell equivalence, i.e. one-to-one map between the spaces of solutions of field equations. Off-shell behaviour of the functionals (\ref{action}) with (\ref{f-GR}) and the Einstein-Hilbert functionals is very different. For a more detailed discussion of the equivalence of the two theories see \cite{Krasnov:2011pp}.

Let us also discuss the relation in the opposite direction. Here one takes an Einstein metric and then computes the self-dual two-forms $\Sigma^i$ given by (\ref{Sigmas}), which satisfy $B^i\wedge B^j\sim \delta^{ij}$. One also computes the self-dual part $A^i$ of the spin connection. It can then be seen that these objects satisfy $d_A B^i=0$, as well as an equation $F^i =\Psi^{ij} B^j$, where $\Psi^{ij}$ are arbitrary coefficients, with the only condition being that ${\rm Tr}(\Psi)\sim \Lambda$. This shows that the self-dual part of the Levi-Civita connection is our sought $A^i$, which satisfies all the relevant field equations. The relation between the two descriptions can be made much more explicit, but here we will refrain from going into more details. Some more details will be given in the next section where we discuss the quantum theory and, in particular, the graviton scattering. 

\subsection{Deformations of GR}

Let us now discuss what happens when the function $f$ used in the construction of the Lagrangian is different from (\ref{f-GR}). As we have already mentioned, when $f(X)={\rm Tr}(X)$ we get a theory without any propagating degrees of freedom. As we have also seen above via the linearization, for any generic choice of $f$ one gets a theory describing a massless spin two particle. Thus, choices of $f$ different from (\ref{f-GR}) give rise to interacting theories of massless spin two particles that are different from General Relativity. Thus, it appears that for a generic choice of $f$ the action (\ref{action}) with $G={\rm SO}(3)$ gives rise to a counterexample to the GR uniqueness theorems, which are often quoted as saying that GR is the only consistent theory of interacting massless spin 2 particles, see e.g. the very first paragraph of \cite{Dvali:2010ue}. The purpose of this subsection is to explain why the above construction of the gauge-theory gravitational theories in no way contradicts the known uniqueness theorems. 

Let us start with a brief review of the GR uniqueness results. This review cannot be exhaustive, since unavoidably only the results known to the author will be mentioned. Historically, the first type of uniqueness results is based on an old idea that the non-linearities of gravity can be reconstructed from the fact that the gravitons must couple (in a canonical way) to their own stress-energy tensor. This line of reasoning is best-known from the description given in \cite{Feynman:1996kb}. Another well-known uniqueness result is \cite{Hojman:1976vp}, which works at the Hamiltonian level and analyses the algebra of constraints of GR. Another approach to GR uniqueness is based on the graviton scattering amplitude kinematical constraints \cite{Grisaru:1975bx}. Finally, the last wel-known approach that we mention is due to Wald \cite{Wald:1986bj}, and analyses the conditions for the infinitesimal gauge symmetries visible at the linearized level to integrate into a gauge symmetry of the non-linear theory. 

We now discuss in turn each of the methods for proving the GR uniqueness. Let us start with the derivations that use the fact that the graviton field $h_{\mu\nu}$ couples to its own stress-energy tensor. In more details, the idea here is that knowing that the spin two field $h_{\mu\nu}$ couples canonically to its own stress-energy tensor (which is computed from the linearised action) one can hope to reconstruct the full non-linear action of gravity. This was achieved (to some extent) in \cite{Feynman:1996kb}, and a much cleaner argument using the first-order formalism was given by Deser in \cite{Deser:1969wk}. A similar argument fixes the coupling of any type of matter to gravity. We note that even in this case there are some controversies and ambuguities, as was recently emphasised in \cite{Padmanabhan:2004xk}. Our next comment is that this type of derivations assume that the only way to describe a spin 2 particle is via a field $h_{\mu\nu}$ with a symmetric pair of spacetime indices. As we have seen, this is not at all the case, and it is equally possible to describe spin 2 particles using a connection variable $a_\mu^i$, where $i=1,2,3$ is an ${\rm SO}(3)$ index. And the proofs based on the $h_{\mu\nu}$ graviton descriptions are simply inapplicable to the connection description. 

One could then ask a question if a similar, not same, logic can be applied in the connection case. However, in this case we are unable to say that the connection $a_\mu^i$ couples to its own stress-energy tensor (as could be derived from the linearised Lagrangian). Indeed, there is simply no notion of the stress-energy tensor in our context (as there is no metric, and so, no notion of the variation of the action with respect to the metric). The best one could hope for is that the connection $a_\mu^i$ couples to some current $J_\mu^i$ that can be obtained (in some canonical way) from the linearised Lagrangian. One could then iteratively determine the non-linearities and completely fix the theory. However, we do not see any physical requirement for why the connection should couple to a certain canonical current, at least in the presently considered case of pure gravity with no matter fields. Indeed, in the case of the usual gravity the requirement of coupling of $h_{\mu\nu}$ to its own stress-energy comes about because one knows how gravity couples to the other matter. The consistency of the field equations (at next to linear order) then require the gravitational stress-energy coupling as well. No similar argument can be made in our present context of pure gravity. The case of gravity plus matter will be commented on below when we discuss gravity plus Yang-Mills unification in our framework. Thus, the line of reasoning based on \cite{Feynman:1996kb}, \cite{Deser:1969wk} is not applicable directly, because a different field is used to describe the spin 2 particles ($a_\mu^i$ instead of $h_{\mu\nu}$). At the same time an analogous argument would require some notion of the universal current $J_\mu^i$ for gravitons, and it is not clear what physical argument would guarantee its existence. 

A more directly applicable line of thought \cite{Wald:1986bj} makes no assumption that gravitons couple to their own stress-tensor. Instead, the idea is simply to analyse the possible non-linear completions of some linear field equations. Then, if some gauge symmetry was present in the linear theory (and thus there was an associated Bianchi identity), then there should be a non-linear version of the Bianchi identity (consistency condition). This consistency condition is then very restrictive and limits the types of possible gauge symmetries. In our case this type of analysis can be applied directly, taking into account the fact that the spin 2 particle is now described by $a_\mu^i$ instead of $h_{\mu\nu}$. One can then see quite easily that there are two types of Bianchi identities satisfied by the field equations (\ref{feqs}). One of these is a differential identity
\be\label{diff-Bianchi}
d_A d_A \left( \frac{\partial f}{\partial X^{ij}} F^j\right)= 0.
\ee
Using the fact that the commutator of two covariant derivatives is the curvature the above identity can be written as
\be
0=\epsilon^{ikl} F^k\wedge \frac{\partial f}{\partial X^{lj}} F^j \sim \epsilon^{ikl} X^{kj}  \frac{\partial f}{\partial X^{lj}},
\ee
where we have used the definition of the matrix $X^{ij}$ as the wedge product of two curvatures. The identity is then a direct consequence of the gauge-invariance of the function $f$. Note that (\ref{diff-Bianchi}) is in fact 3 equations, and is a differential identity for the field equations. It is clear that this identity has its origin in the gauge invariance of the action.

Another type of identity does not involve any derivatives of the field equations. It reads:
\be\label{Bianchi}
\iota_\xi F^i \wedge d_A \left( \frac{\partial f}{\partial X^{ij}} F^j\right) = 0,
\ee
and, as we shall see, holds for any vector field $\xi$. Here $\iota_\xi$ is the interior multiplication with a vector field $\xi$ (and, in (\ref{Bianchi}), the curvature 2-form). To prove this identity we use the "usual" Bianchi identity $d_A F^i=0$ to take the curvature 2-form $F^j$ out of the brackets. The quantity $\iota_\xi F^{(i} \wedge F^{j)}$, where the brackets denote the symmetrization, is then proportional to $\iota_\xi ({\rm vol}) X^{ij}$, where $({\rm vol})$ is the volume form used to define $X^{ij}$ via $F^i\wedge F^j = ({\rm vol}) X^{ij}$. So, the above identity reduces to 
\be
X^{ij} d_A  \left( \frac{\partial f}{\partial X^{ij}} \right) = 0,
\ee
which is a consequence of the homogeneity (of degree one) of the function $f$. It is clear that (\ref{Bianchi}) is related to the property of the action being invariant under the diffeomorphisms. We also note that the above arguments are valid for any choice of the gauge group, provided $\epsilon^{ijk}$ in the proof of the first identity is replaced with the appropriate structure constant. 

To summarise, we see that for any choice of $f$ the theory (\ref{action}) satisfies two Bianchi-type identities, with one of them (\ref{diff-Bianchi}) being a differential identity involving the field equations, and the other (\ref{Bianchi}) being simply a statement that the field equations are not linearly independent. The gauge symmetries of the theory are then the usual gauge rotations as well as diffeomorphisms. No additional requirements is placed on the allowed interactions. In particular, no restrictions are placed on the type of function $f$ that can be used in construction of the Lagrangian, provided this function satisfies the requirements of gauge invariance and homogeneity. No unique Lagrangian (such as that of GR) can follow from such arguments. 

Another GR uniqueness claim \cite{Hojman:1976vp} is based on the canonical analysis. The main idea of the argument is that the phase space of any theory of gravity is that of pairs (spatial metric, conjugate momentum). One also knows that the spatial metric must appear in the result of the commutator of two Hamiltonian constraints. This, together with some assumptions, allows one to fix the form of the Hamiltonian constrain and thus the gravitational dynamics. This argument is not applicable in our case because it can be shown by a direct analysis, see \cite{Krasnov:2007cq}, that the canonical variable that arises in the case of theories under consideration is not the spatial metric. The algebra of diffeomorphisms is indeed the usual expected algebra, but in our case the spatial metric appearing in the commutator of two Hamiltonian constraints turns out to be a complicated function of the canonical variables. It thus appears that the form of the Hamiltonian cannot be fixed via this type of argument in the case when the configurational canonical variable is not directly related to the (spatial) metric. 

Finally, yet another argument for GR uniqueness \cite{Grisaru:1975bx} is based on the analysis of the 4-graviton scattering amplitude and showing that it is essentially determined (up to a numerical factor) by the kinematical constraints only. However, this analysis assumes parity invariance of the amplitudes (to reduce the number of the independent amplitudes). This assumption is explicitly violated by a generic theory from the our family, with only the theory with (\ref{f-GR}) being parity invariant. Thus, while it appears that there is indeed a unique parity invariant gravity (which is GR), there exists also a large family of interacting theories of massless spin 2 particles that are not invariant under parity. 

Much more can be said about the classical behavior of the deformations of GR, but we cannot go into such details because of the space restrictions. We just mention that the spherically symmetric (BH) solution can be obtained for an arbitrary $f$, see \cite{Krasnov:2007ky} and also \cite{Krasnov:2008sb} for a more informal discussion. We note that the reference \cite{Krasnov:2007ky} uses a certain equivalent (but not obviously so) formulation of the same class of theories (similar to Plebanski formulation of GR). A derivation of the spherically-symmetric solution is also possible directly in the "pure connection" framework presented here, but this will be spelled out elsewhere. We note in passing that a general theory from our family exhibits an interesting resolution of the singularity inside the black hole phenomenon, which is described in details in \cite{Krasnov:2007ky}.

\subsection{Gravity-Yang-Mills unification}

Our above discussion of the coupling of gravitons to themselves (and to other matter) has already made it clear that this coupling is not as straightforward as in the usual metric-based description. In particular, there seems to be no notion of some canonical stress-energy tensor via which all matter would couple (and gravitons would self-couple). It then appears that the only consistent way to couple matter to our gravity theory is by enlarging the gauge group in the action (\ref{action}). As we shall soon see, an ${\rm SO}(3)$ subgroup of such a larger group will still describe gravity, while the part that commutes with this ${\rm SO}(3)$ will describe the Yang-Mills sector. Constructions of this sort where first spelled out (in the framework of Plebanski-like formulation) in \cite{TorresGomez:2009gs}, \cite{TorresGomez:2010cd}. For earlier work with similar aims (in the context of the Hamiltonian formulation) see e.g. \cite{Chakraborty:1994vx}. Here we describe what appears to be a much simpler derivation. 

Thus, let us take a general theory from the family (\ref{action}), with some gauge group $G$. As we have already discussed above, a particularly nice set of backgrounds, or "vacua" for our theory is provided by (\ref{A-backgr}). Such a background connection selects some ${\rm SU}(2)\sim{\rm SO}(3)$ subgroup of $G$. Note that there are in general several inequivalent embeddings of ${\rm SU}(2)$ into a larger gauge group. The idea is to fix such an embedding and look at the spectrum of excitations that arise around this background. Let us first look at the linearised theory, i.e. an expansion of the action up to second order in the fields. It is then easy to see that the ${\rm SU}(2)$ part of the gauge field perturbation is still described by the linearised action (\ref{S-lin-gr}), as no details of the derivation presented in the corresponding subsection get changed. So, we still get two propagating polarisations of the graviton in this sector. 

Let us now discuss the sector of the theory described by $\delta A^a$ with the index $a$ being in the part of the Lie algebra of $G$ that commutes with the fixed ${\rm SU}(2)$. We note that this part may be empty (this depends on the embedding chosen). Let us first consider the terms in the second line of (\ref{sec-var}). The first term becomes (after the limit $M\to 0$ is taken) a total derivative and can be dropped. In the second term, for the background chosen, the only non-zero values of $F^a$ are in the ${\rm SU}(2)$ part of the Lie algebra. But by assumption, the commutator of the two connections in the part of the Lie algebra that commutes with ${\rm SU}(2)$ does not have any component in the ${\rm SU}(2)$ part. So, the second term in the second line of (\ref{sec-var}) is also zero. We are left only with the first line. Here the matrix of the second derivatives of the function $f$ must be evaluated on the background $X^{ab}\sim (\delta^{ij}, 0)$, where one has $\delta^{ij}$ in the ${\rm SU}(2)$ part of the matrix and zero everywhere else. Also, one is only interested in $\partial^2 f/\partial X^{ia}\partial X^{jb}$ part of the matrix of the second derivatives. One can easily convince oneself that this matrix can only be proportional to $\delta^{ij} \delta^{ab}$. Thus, we can write:
\be\label{f-2-ym}
\frac{\partial^2 f}{\partial X^{ia}\partial X^{jb}} = \frac{\kappa}{2} \delta^{ij} \delta^{ab},
\ee
where $\kappa$ is some constant. Note that this holds only for the part of the Lie algebra that commutes with the ${\rm SU}(2)$, because for the part that does not commute there are other possible invariants that can appear on the right hand-side of (\ref{f-2-ym}). Collecting all the constants (and dividing the second variation by two), we get the following linearised Lagrangian:
\be\label{S-lin-YM}
S^{(2)}_{\rm YM} = -\frac{\kappa}{2} \int \delta^{ij}\delta^{ab} (\Sigma^{i\mu\nu}\partial_\mu \delta A^a_\nu) (\Sigma^{i\rho\sigma}\partial_\rho \delta A^b_\sigma),
\ee
where we already took the limit $M\to 0$. As in the ${\rm SU}(2)$ sector, the factors of $M$ have cancelled out in this derivation, and the only effect of the limit is in replacing the covariant derivatives by the usual ones (and in removing the $\sqrt{-g}$ factor). We now rescale $a^a_\mu = \sqrt{\kappa} \delta A^a_\mu$ and get the usual linearised Yang-Mills action in the form (\ref{L-YM-S}) encountered in the beginning of this paper. 

Thus, it is very easy to see how both gravity and Yang-Mills linearised theories arise from the "mother" theory (\ref{action}), once it is expanded around the background (\ref{A-backgr}). We see that the first of the expectations spelled out in the Introduction, namely that a gauge-theoretic reformulation of gravity may be of help for the problem of unification with the other forces, seems to be at least partially fulfilled. 

\subsection{Symmetry breaking}

As we described above, backgrounds (\ref{A-backgr}) that can be used as "vacua" for our theories are in one-to-one correspondence with embeddings of the Lie algebra ${\mathfrak sl}(2)$ into the full Lie algebra $\mathfrak g$. We have discussed the interpretation of the connection components charged under the part of the Lie algebra $\mathfrak g$ that commutes with the embedded ${\mathfrak sl}(2)$. Thus, we have seen that these describe massless gauge bosons. It can be shown \cite{Krasnov:2011hi} that the other connection components, i.e. in those directions in $\mathfrak g$ that do not commute with the embedded ${\mathfrak sl}(2)$ typically describe massive fields of non-zero spin. 

A very interesting symmetry breaking scenario then becomes available. One can change the "vacuum" around which the theory is studied. In one vacuum, where the centralizer of the embedded ${\mathfrak sl}(2)$ is non-trivial we have the corresponding massless gauge bosons. In a vacuum with trivial centralizer all massless gauge bosons of the previous vacuum become massive particles. Thus, by choosing the background connection (and thus the embedding of ${\mathfrak sl}(2)$) appropriately, we can break the Yang-Mills gauge group as desired. 

This symmetry breaking mechanism is described in details in \cite{Krasnov:2011hi}. Here we restrict ourselves to just illustrating how this mechanism works on the example of $G={\rm SL}(3)$. Note that we work with complex Lie groups, so we make no distinction between e.g. ${\rm SU}(3)$ and ${\rm SL}(3)$. There are two inequivalent embeddings of ${\rm SL}(2)$ into ${\rm SL}(3)$. One is the obvious one that embedds $2\times 2$ matrices in e.g. the upper left corner of $3\times3$ matrices in ${\rm SL}(3)$. This embedding breaks the original symmetry down to ${\rm SL}(2)\times {\rm U}(1)$, where the latter is generated by the matrices ${\rm diag}(1,1,-2)$. As we have seen in the previous subsection, components of the connection charged under this ${\rm U}(1)$ gauge group describe massless gauge bosons - i.e. a Maxwell electromagnetic potential. The other fields that arise in this background turn out to be massive spin $3/2$ particles that are electrically charged. There are two such fields, oppositely charged. It is unusual to have fields of half-integer spin described by commuting connection components, and so one might worry about the Hamiltonian being not bounded from below. However, the Lagrangian for this fields is second order in derivatives (and is of the same type as all $\Sigma$-containing Lagrangians we have encountered above). It is not hard to choose the reality conditions so that the Hamiltonian is explicitly positive-definite, and so no problems of the type seemingly predicted by the spin-statistics theorem arise. For more discussion on these issues the reader is directed to \cite{Krasnov:2011hi}.

Let us now consider the other inequivalent embedding. In this case there is no non-trivial centralizer of the embedded ${\rm SL}(2)$ in ${\rm SL}(3)$, and all the gauge symmetry apart from the gravitational ${\rm SL}(2)$ is broken. The massive field content of this model is as follows. One finds a massive spin 3, as well as a massive spin one field. We can say that the massless gauge boson have absorbed one of the components of the massive fields of the previous background, and became massive. The other massive fields have rearranged themselves into a spin 3 field. 

To summarize, we have all the ingredients of a symmetry breaking mechanism, where changing the background we can break the gauge symmetry as desired, with massless gauge bosons of one background becoming massive in another. Many more examples of this mechanism are described in \cite{Krasnov:2011hi}, including those with a realistic Standard Model symmetry breaking pattern. We refer the reader to this reference for details. 

We finish this section by stating that, in spite of the fact that our description of gravity is so very far from the standard one that there is a legitimate worry that one will never be able to couple any realistic matter to it, at least some types of matter can be coupled very naturally. Thus, we have seen that the gauge fields can be coupled just by enlarging the gauge group. Quite interestingly, massive particles can also be coupled in the same way. In particular, and this is far from obvious, spin zero massive particle can also arise in this framework, see \cite{Krasnov:2011hi} for an example. It remains to be seen whether the Standard Model fermions can be embedded into this framework. It currently seems that if this is to be possible, the anti-commuting fermionic variables must be described by the grading-odd elements of some super Lie-algebra.

\section{Quantum theory}
\label{sec:quant}

In this section we finally discuss whether our gauge theoretic formulation of gravity sheds any new light on the problem of quantum gravity. We start with a description of how the perturbative quantum theory looks like in the new language.

\subsection{Interactions}

Like in the metric-based General Relativity, one can expand the full non-linear action (\ref{action}) around some background, and then attempt to compute quantum corrections to the theory by evaluating Feynman diagrams. If one wants the asymptotic states to exist (one usually does) one has to work around the Minkowski spacetime background. One complication (that can be overcome successfully) in our case is that the Minkowski spacetime connection is $A=0$, and as such it is difficult to expand around (imagine having to expand the Einstein-Hilbert action around the zero metric). So, the natural strategy appears to take the constant curvature background, and expand the action around it, taking the flat limit afterwards. We have already seen how the linearised gravity action (\ref{S-lin-gr}) appears this way. One can continue this process, and compute the interactions vertices. It turns out, however, that one should be very careful about taking the limit $M\to 0$. Indeed, one finds e.g. that the cubic interaction vertex for our gravitons is of the schematic form:
\be\label{L3}
{\cal L}^{(3)}\sim \frac{1}{M M_p} (\partial a)^3 + \frac{M}{M_p} (\partial a) a^2.
\ee
Here we have considered the case of the theory with $f$ given by (\ref{f-GR}) that corresponds to General Relativity. Then $M_p$ is the Planck mass $M_p^2\sim 1/G$. For a general $f$ the cubic interaction is schematically the same, except that one has $1/M^2$ in front of the first term, multiplied by some dimensionless combination of the parameters of $f$, see below. We remind the reader that the parameter $M$ is the length scale introduced by the background (\ref{F-backgr}), which is related to the cosmological constant as in (\ref{M-Lambda}). For details of this derivation the reader can consult \cite{Krasnov:2011up} (or perform a straightforward analysis him/herself). 

There are several points to be noted about (\ref{L3}). First, there are non-renormalisable (by power-counting) interactions, such as the first term in (\ref{L3}). Second, the interactions seem to blow up in the limit $M\to 0$. Finally, unlike in the metric based GR, we get more than two derivatives in the interaction vertices. This can be easily seen to be a general feature, and the order $n$ vertex will have as many as $n$ derivatives in the form $(\partial a)^n$. We can write:
\be\label{Ln}
{\cal L}^{(n)}\sim \frac{1}{(M M_p)^{n-2}} (\partial a)^n + \ldots,
\ee
where the dots denote terms that have a lower number of derivatives. We also note that the number of derivatives in each successive term changes by two (as in (\ref{L3})), so the next term in (\ref{Ln}) is proportional to $(\partial a)^{n-2} a^2$. Thus, the perturbation theory is seemingly very different from the standard one. 

Both of these puzzles (blowing up interactions and the higher number of derivatives) can be understood by noting that there is a relation between the connections and the metric perturbations, which is valid on-shell only, and which reads, schematically,
\be\label{h-a}
h = \frac{1}{M} \partial a.
\ee
There is yet another relation which is also valid on-shell, and which reads:
\be\label{a-h}
a = \frac{1}{M} \partial h.
\ee
Both of these can be true at the same time only on-shell $\partial^2 = M^2$. These relations can be used, in particular, to determine the helicity spinors in the connection description from those in the metric description \footnote{The helicity spinors can of course be built independently from any metric description, but the two ways of determining the helicity states agree.}.

We see that it is tricky to take the limit $M\to 0$. This certainly cannot be done at the very start of the computation, because the interaction vertices seem to diverge in this limit, and so do the helicity states determined from e.g. (\ref{a-h}). Thus, what one has to do at intermediate stages of the calculation is to work  on the mass-shell $\partial^2 = M^2$, and then only take the limit $M\to 0$ after the physical quantities (such as the graviton scattering amplitudes) are computed. This procedure can be shown to lead to the well-known graviton scattering amplitude results \cite{DKS}. 

Let us describe in a bit more detail how e.g. the 4-graviton scattering amplitude can be reproduced. As in the metric-based GR one can show that the 4-valent interaction vertex cannot contribute to the answer. Thus, one has to analyse only the contribution from diagrams involving two 3-valent vertices. The 3-valent interaction vertex is of the schematic form (\ref{L3}). However, let us make the structure of the expression in (\ref{L3}) more explicit. The spinor formalism is again most suited for this purpose. So, we convert all spacetime as well as internal indices into spinor ones. Recall that the connection becomes an object $a^{ABCA'}$ with three unprimed and one primed spinor index. The relation (\ref{h-a}) then reads:
\be\label{h-a-1}
h_{AB A'B'} = \frac{1}{M} (\partial a)_{AB A'B'},
\ee
where we do not worry about constants (which are in any case convention dependent). We used the following notation for the spinor contractions:
\be\label{da}
(\partial a)_{AB A'B'}:=\partial^E_{(A'} a_{B')EAB}, \qquad (\partial a)_{ABCD}:= \partial_{E'(A} a^{E'}_{BCD)}.
\ee
The second of these is to be used in the cubic interaction vertex below. In both expressions here we assume the connection on the right-hand-sides to take values in $S_+^3\otimes S_-$, i.e. be totally symmetric in its 3 unprimed indices. As we have discussed in the section on linearised theory, this is the way to project out the components of the connection that are pure (diffeomorphism) gauge. The object on the left-hand-side in (\ref{h-a-1}) is then $AB$ and $A'B'$ symmetric, and is the spinor representation of the tracefree part of the graviton field $h_{\mu\nu}$. 

Let us also give the spinor expression for the 3-valent interaction vertex that one gets for the case of $f$ corresponding to GR. Using the notations (\ref{da}) we have:
\be\label{L3-1}
{\cal L}^{(3)}\sim \frac{1}{M M_p} (\partial a)^{ABCD} (\partial a)_{ABE'F'} (\partial a)_{CD}{}^{E'F'} + \ldots,
\ee
where the dots denote terms that go to zero in the limit $M\to 0$. Using this vertex (and appropriate spinor helicity states for the external gravitons) one can reproduce many of the GR graviton scattering amplitudes. In particular, the usual 4-graviton scattering amplitude can be obtained, as well as the general MHV amplitude (it can be shown that only this vertex contributes to the MHV amplitudes). The later is obtained using essentially the same Berends-Giele-type recursion relations \cite{Berends:1987me} that work in the case of MHV amplitudes of gluon scattering in Yang-Mills theory. Details of these computations will appear in \cite{DKS}. 

Before we compare the above vertex to that in the case of metric-based GR let as note that it is analogous to the cubic vertex of Yang-Mills theory (with the Lie algebra index suppressed):
\be\label{L3-YM}
{\cal L}^{(3)}_{\rm YM} \sim g_{\rm YM} (\partial a)^{AB} a_{AE'} a_B{}^{E'},
\ee
where $(\partial a)_{AB}:=\partial_{E'(A} a_{B)}{}^{E'}$ is essentially the self-dual part of the curvature, and $g_{\rm YM}$ is the Yang-Mills coupling constant. The main difference between our gravity vertex (\ref{L3-1}) and the above Yang-Mills one is that the latter is renormalizable while the former is not. This is related to two extra derivatives being present in (\ref{L3-1}) as compared to (\ref{L3-YM}). The point that we would like to emphasise however is that the gauge theory formulation graviton interaction vertex is as simple as that of Yang-Mills theory, which gives hope that computational complexity can be greatly reduced using this formalism. This hope is indeed realised, and we again refer the reader to the upcoming work \cite{DKS}. 

Let us now quickly relate the vertex (\ref{L3-1}) to the cubic vertex in the metric-based formalism. We will only show the schematic form, because even the translation of (\ref{L3-1}) to the metric variables using the on-shell relations (\ref{h-a}) and (\ref{a-h}) is rather involved. We get:
\be
{\cal L}^{(3)}\sim \frac{1}{M_p} (\partial^2 h) hh,
\ee
which is the usual 3-valent vertex of the metric based GR. In short, this is the explanation how such a seemingly different theory as the one with interactions (\ref{L3-1}) can reproduce the usual graviton scattering amplitudes. 

Further, it is interesting to note that in the connection description the parity invariance of General Relativity becomes not manifest. The parity invariance can be shown to hold (at least order by order in perturbation theory) for the theory (\ref{f-GR}). But it can also be seen to be explicitly violated by any other general member of the family (\ref{action}). Thus, e.g. for the 4-graviton scattering in a generic theory one finds that  the all minus and all minus one plus amplitudes are zero, while the all plus and all plus one minus amplitudes are not. In the usual case only the MHV amplitude $(++--)$ is non-zero for the 4-graviton scattering. So, the general member of the deformations of GR family can be seen to violate parity quite explicitly. 

In preparation for a discussion of the quantum theory, let us briefly describe the structure of the interactions for a general $f$. In the case of $f$ corresponding to GR there are certain delicate cancelations that guarantee that there is no ${\rm Tr}((\partial a)_{ABCD})^3$ term in (\ref{L3-1}). For a general function $f$ there is no such cancelation, and the structure of the 3-vertex is
\be\label{L3-2}
{\cal L}^{(3)}\sim \frac{c_1}{M^2} (\partial a)_{AB}{}^{CD} (\partial a)_{CD}{}^{EF} (\partial a)_{EF}{}^{AB} 
\\ \nonumber 
+ \frac{c_2}{M^2} (\partial a)^{ABCD} (\partial a)_{ABE'F'} (\partial a)_{CD}{}^{E'F'} + \ldots.
\ee
Note that the quantity $M_p$ no longer appears, as there is no notion of the Planck mass for a general theory from the family (\ref{action}). Indeed, recall that there are only dimensionless parameters in any of our theories (\ref{action}). Thus, the quantities $c_1, c_2$ above are dimensionless parameters related to the derivatives of the function $f$ evaluated at the identity matrix, and $M$, as before, is the length scale introduced by the background. For the $f$ corresponding to GR we have $c_1=0, c_2\sim M/M_p$, but for a general theory these are arbitrary dimensionless parameters. The relations (\ref{h-a-1}) and (\ref{a-h}) are still valid, and so the leading term of the 3-vertex can be rewritten in the metric terms as
\be\label{L3-3}
{\cal L}^{(3)}\sim  \frac{c_1}{M^5} (\partial^2 h)^3 + \frac{c_2}{M} (\partial^2 h) hh + \ldots.
\ee
The point of this exercise is to exhibit explicitly that the interaction vertices of a general theory from the family (\ref{action}) contain, when rewritten in the metric language, terms with more than two derivatives. In general, the n-valent vertex will start from a term $(\partial^2 h)^n\sim (Riemann)^n$ when rewritten in the metric language. This fact will be important below when we discuss how the theories (\ref{action}) behave under the renormalisation.

\subsection{Renormalisation}

So far we have only talked about the tree-level graviton scattering amplitudes. These can be expected to agree with the results obtained in the usual metric-based approach (for the theory (\ref{f-GR})). However, as we have already discussed, the theory (\ref{f-GR}) is only on-shell equivalent to that described by the Einstein-Hilbert Lagrangian. Off-shell the two actions have very different properties. Thus, one can expect that the quantum theory based on (\ref{f-GR}) is different from that arising in the metric approach. More specifically, one should not expect the loop-corrected graviton scattering amplitudes to be the same in the two approaches. 

With this general remarks being made, let us discuss the issue of non-renormalisability of GR. In the usual metric formalism it manifests itself on-shell starting at the two-loop order, see \cite{Goroff:1985th}. At this order of perturbation theory one finds it necessary to add a new $(Riemann)^3$ term to the Einstein-Hilbert Lagrangian to cancel the arising divergence. This is expected to continue at higher-loop orders (even though nobody has calculated the 3-loop divergences in pure gravity explicitly). It is thus believed that an infinite number of higher derivative terms has to be added to the Einstein-Hilbert Lagrangian in the process of its renormalisation. This introduces an infinite number of new coupling constants, all becoming equally important around the Planck scale, and implying loss of predictive power of the theory at Planckian energies. 

Let us discuss how the non-renormalisability of gravity can manifest itself in the gauge-theoretic description. We have seen that (some of) the coupling constants of our theory, once the Lagrangian is expanded around a constant curvature background, are of negative mass dimension, as e.g. in the first term in (\ref{L3}). It is thus to be expected that higher derivative counter-terms will get produced in the process of renormalisation of the loop corrections. However, unlike the case with the metric-based GR, where the perturbative expansion of the Lagrangian only contains two-derivative vertices, the vertices in our description already contain arbitrarily high powers of the derivative, see (\ref{Ln}). This fact will be very important below. 

Before we discuss implications of the higher-derivative nature of our perturbative expansion, we emphasise that the appearance of all powers of the derivative does not mean that the theory is non-local. Indeed, as we have discussed in the previous section, the Lagrangian of the theory is non-linear, but leads to not higher than second order in derivatives field equations. So, there is no non-locality in the theory, but, rather just  non-linearity that produces arbitrarily high powers of $(\partial a)$ when the Lagrangian is expanded. 

The main point is then that, since arbitrary high powers of the derivatives are already present in the expanded Lagrangian, the higher-derivative terms that get produced in the process of the renormalisation may already be contained in the original Lagrangian. Indeed, we have already seen how the 3-vertex (\ref{L3-3}) of the general theory, when rewritten in the metric language starts from the term $(\partial^2 h)^3$, which is nothing else but $(Riemann)^3$. This is exactly the counterterm that is known \cite{Goroff:1985th} to be needed in the usual metric based quantum gravity at two loops. Thus, at least to the lowest order the idea that the terms that have to be added to our Lagrangian in the process of renormalisation are already contained in it seems to hold.  It is far from being trivial that this continues to happen to all orders, but this can in principle happen, as it contradicts nothing we know about the behavior of non-renormalisable theories. The important point here is not to see that the even the higher-derivative terms appear, for we have seen that the general $n$-valent vertex will start from the term $(Riemann)^n$. What would be non-trivial is that it is precisely the structure that can be encoded into a general Lagrangian of the form (\ref{action}) that gets preserved by the process of the renormalisation. For a more explicit calculation showing that the $(Riemann)^3$ term that is known to arise in usual gravity at two-loops does appear in the perturbative expansion of a general Lagrangian from the family (\ref{action}) we refer the reader to \cite{Krasnov:2009ik}.

To summarise, thanks to the peculiar higher-derivative structure of the perturbatively expanded Lagrangian of our theory, it may be that the class of theories (\ref{action}) is closed under the renormalisation. As we have already said, this is far from being trivial, for we know that the class (\ref{action}) is not the most general class of diffeomorphism invariant gauge theories. Indeed, as we have discussed in the previous section, the theories that lead to higher than second order field equations are not included into (\ref{action}). We have then emphasised that "not higher than second order field equations" is not equivalent to "not higher than two derivatives in the interaction vertices", for the non-linearities of a second-derivative field equations Lagrangian can (and do) manifest themselves as higher-derivative interactions. The conjecture is then that in the process of renormalisation of (\ref{action}) only similar non-linearities appear, but no terms that change the order of the field equations. 

This conjecture was first stated in \cite{Krasnov:2006du} in a different (Plebanski-like formulation) of this class of theories. It is certainly non-trivial to prove, and the ongoing attempt is to compute the one-loop correction to the general Lagrangian (\ref{action}) using the background field method. If it is found that at least at one-loop one does not produce anything that is not already contained in (\ref{action}), this would give a partial support to the conjecture. In the next subsection we shall discuss a possible scenario for the UV behavior of gravity in case the conjecture discussed in this subsection is correct. 

\subsection{Conjectural RG flow}

In this subsection we shall assume that the class of theories (\ref{action}) is closed under the renormalisation and discuss what this implies for the problem of finding the UV completion of gravity. The closedness under the renormalisation can be expressed as a property that the beta-function governing the renormalisation group flow of the Lagrangian (\ref{action}) is of the same form as the Lagrangian itself:
\be\label{RG}
\frac{\partial f(F\wedge F)}{\partial \log{\mu}} = \beta_f(F\wedge F).
\ee
Here $\beta_f(F\wedge F)$ is a function with the same properties as $f$, i.e. a gauge-invariant, and homogeneous function of degree one of a symemtric $3\times 3$ matrix, which can thus be applied to a matrix-valued 4-form. The subscript $f$ indicates that this function depends on the function $f$ itself (i.e. the beta-function is a function of the coupling constants encoded in $f$).

The ultra-violet behavior of the theories (\ref{action}) is then controlled by the UV fixed point(s) of the RG flow (\ref{RG}). One is then in the domain of applicability of the asymptotic safety ideas, see e.g. \cite{Weinberg:2009bg} for a recent description. The idea of the asymptotic safety is that there is a UV fixed point of the relevant RG flow (such as e.g. (\ref{RG})), with only a finite number of attractive directions. In other words, the idea is that most of the RG trajectories going to the UV approach the relevant fixed point, but then miss it and escape to infinity in the space of theories (coupling constants). And only a finite-dimensional subset of the trajectories actually end at the fixed point. The fixed point in question then provides the UV completion of the theory, and one gets predictive power from the requirement that the physical theory is on one of the trajectories that leads to this fixed point. Then, since the dimension of the surface of such trajectories is finite, one needs only a finite number of measurements to fix the theory. 

If the RG flow for the class (\ref{action}) is of the form (\ref{RG}), the ideas of asymptotic safety can be applied directly. The flow is then essentially just a flow in the space of functions of two-variables $f(X)=f(\lambda_1,\lambda_2,\lambda_3)=\lambda_1 f(1,\lambda_2/\lambda_1,\lambda_3/\lambda_1)$, where we have used the fact that a gauge-invariant function of a symmetric $3\times 3$ matrix $X^{ij}$ is a function of its eigenvalues $\lambda_1,\lambda_2,\lambda_3$, where $X^{ij}=O {\rm diag}(\lambda_1,\lambda_2,\lambda_3) O^T$ and $O\in {\rm SO}(3)$ is the diagonalising orthogonal transformation. Such a flow can be studied very explicitly, and the question of whether there are any UV fixed points with a finite number of attractive directions becomes answerable.

We then note that the ideas of asymptotic safety have a potential of explaining why it is the function (\ref{f-GR}) that appears as the low-energy Lagrangian. Indeed, one could envisage that being on a trajectory that emanates from the UV fixed point necessarily implies that one flows to (\ref{f-GR}) at low energies. This way one could even explain the fact that the dimensionless constant in front of the GR Lagrangian (\ref{f-GR}), namely $1/(16\pi G\Lambda)\sim 10^{120}$, is so large. Indeed, it is not impossible that the large numbers of this sort are produced by the RG flows. Thus, it may be that our scenario has something to say about the very difficult cosmological constant problem, which is to explain why such two different scales seem to co-exist in Nature. 

Returning to the problem of the UV completion of gravity, we note that there is one point in the space of theories (\ref{action}) that should certainly be a fixed point of the flow (\ref{RG}). This is the topological point $f_{top}(X)={\rm Tr}(X)$. Indeed, while this point is not strictly speaking among the class of theories that we consider (dynamically non-trivial theories with 2 propagating degrees of freedom), it should certainly annihilate the beta-function 
\be\label{beta-top}
\beta_{f_{top}}(F\wedge F)=0.
\ee
The reason for this expectation is that a topological theory without any dynamics can have no non-trivial running, and this should imply (\ref{beta-top}). It can be conjectured that the fixed point (\ref{beta-top}) is the sought UV fixed point. If this is the case, then the dynamics in the UV region is controlled by this fixed point, with the "safe" theories approaching this fixed closer and closer as one increases the energy. The arising conjectural picture of the UV behavior of gravity is then depicted in Figure 1.

\begin{figure}
\label{fig}
\begin{center}
\includegraphics[width=90mm]{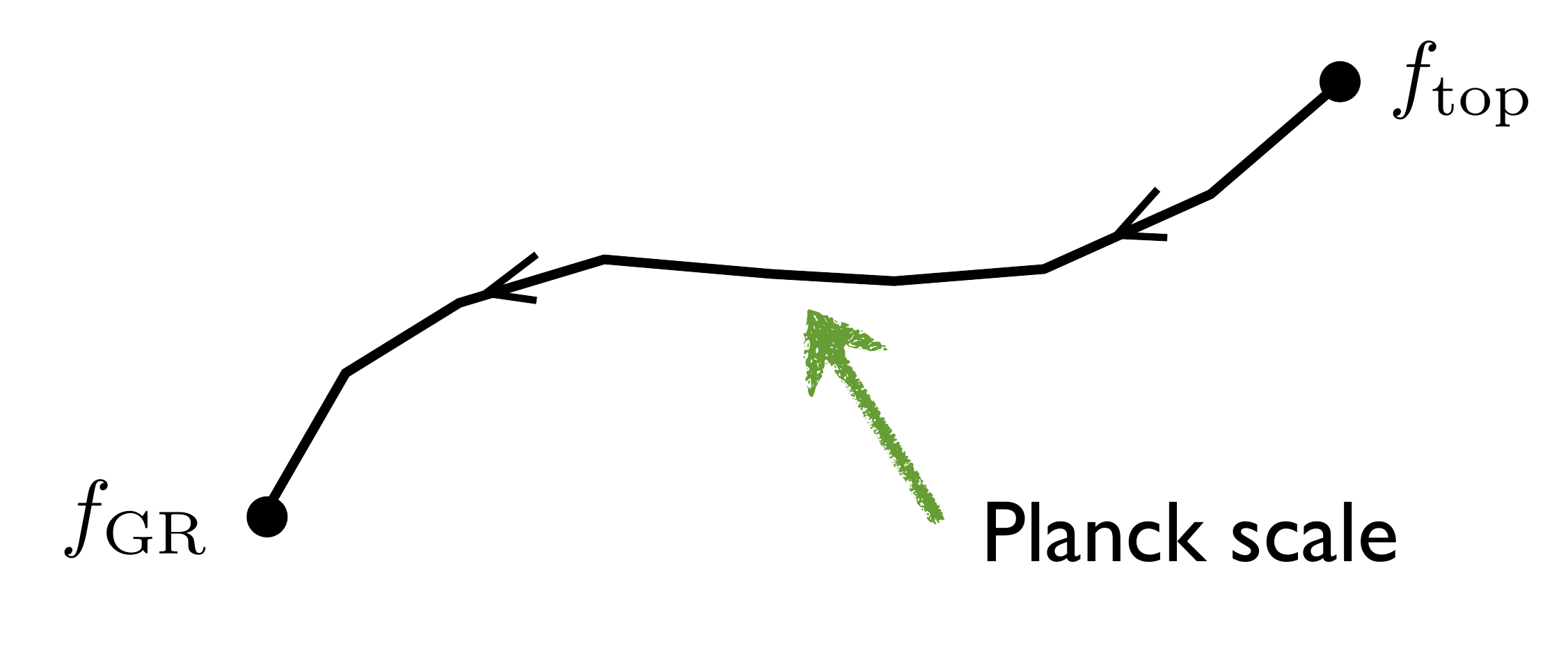}
\caption{The conjectural RG flow}
\end{center}
\end{figure}

\section{Conclusions}
\label{sec:concl}

We have come a long way: from a simple Lagrangian (\ref{L-GR}) (see also (\ref{L-GR-spinor})) that describes free gravitons in Minkowski spacetime, to a possible scenario of the UV completion of quantum gravity. We have also seen how both spin 2 particles (gravitons) and spin 1 particles ("gluons") are simultaneously described by the same theory (\ref{action}), once it is expanded around an appropriate background. Here we would like to briefly reiterate the main ideas of the approach of this paper, and then list questions that remain open.

One of the main ideas was to use a gauge field rather than the metric as the main dynamical variable for gravity (in particular GR). We have seen how this is possible at both the linearised level (\ref{L-GR}), and, more non-trivially, at the full non-linear level via theories (\ref{action}) with (\ref{f-GR}). We have seen that gravity formulated in this language gets naturally embedded into a much larger class of theories (\ref{action}), to which we referred to as diffeomorphism invariant gauge theories. The gravitational sector of such a general theory is then described by an appropriate ${\rm SU}(2)$ subgroup, while the other parts of the gauge group describe matter. In particular, the part of the gauge group that commutes with the gravitational ${\rm SU}(2)$ describing the Yang-Mills field. 

Our construction of the class of theories (\ref{action}) seems a very natural one, as it only uses as the input the principles of diffeomorphism and gauge invariance (which are both known to be of fundamental importance), as well as the desire to have the field equations of not higher than second order in the derivatives. We note that, importantly, there are no dimensionful parameters in our class of theories, with the dimensionful constants only arising once the theory is perturbatively expanded. This feature of theories (\ref{action}) of not having any dimensionful couplings is, in our opinion, a very desirable one for any would be fundamental theory.

Importantly, our theories are non-linear, while still leading to just second order field equations. Non-linearity is of course unavoidable in the real world, for without it there would be no interactions between particles. The most interesting theories that play role in the description of Nature -- Yang-Mills and General Relativity -- are both non-linear. However, the non-linearity of (\ref{action}) is much stronger. In the case of Yang-Mills theory the Lagrangian is polynomial in the fields (there are only up to quartic interaction vertices). In GR non-linearity is worse, because once expanded (around a background) the action of General Relativity contains vertices of arbitrarily high valency. Still, the order of the derivative that appears in GR vertices is always two (or smaller if one also takes into account the vertices coming from the cosmological term). The non-linearity of our theories is similar to that in GR in that interaction vertices of arbitrary valency appear. However, it can be said to be stronger in the sense that up to the $n$-th power of the derivative can appear at the $n$-valent vertex. On shell, there is no contradiction between the two approaches. Indeed, as we have discussed, on-shell the connection variable is related to the derivative of the metric one, see (\ref{a-h}). This explains how what looks like an expansion in powers of $\partial a$ can be rewritten as an expansion in powers of the metric perturbation. However, as we have discussed in details in the first section of this paper, there can be no off-shell relation between the gauge theoretic and the metric descriptions. Therefore, off-shell, and in particular in quantum mechanical computations, one has to take the peculiar form of the perturbative expansion (\ref{Ln}) of the gauge theoretic gravity Lagrangian seriously. It is then that the most interesting prospect, which may lead to an eventual UV completion of these theories, arises. 

Both GR and our gauge theories (\ref{action}) are non-renormalisable. But in GR the non-renormalisability manifests itself in a very drastic way with new terms being added to the Lagrangian at every loop order to cancel the arising divergences. These terms are of higher order in the derivatives, and for this reason they could not have been in the original second derivative Einstein-Hilbert Lagrangian. In the case of theories (\ref{action}) we can also expect UV divergences that will need to be renormalised. However, the principal difference between (\ref{action}) and the GR Lagrangian is that (\ref{action}) already contains arbitrarily high powers of the derivative when perturbatively expanded. This comes from the non-linearity of the theory, and,  as we discussed above, does not contradict the fact that its field equations are second order. It may then be that the higher-derivative counter-terms that are needed for the renormalisation of a theory from the class (\ref{action}) are of the same general form (\ref{action}). In other words, as we have conjectured, it may be that the class (\ref{action}) is closed under the renormalisation. 

We then appealed to the ideas of asymptotic safety and gave a possible scenario for the UV fixed point in this class of theories. Thus, the sought UV fixed point that controls the dynamics of the theory at ultra-high energies may be the topological fixed point, corresponding to a theory without any propagating degrees of freedom. However, whatever the UV fixed point may be, if the idea of the closedness of (\ref{action}) under the renormalisation is correct, the RG flow (\ref{RG}) has a potential of providing a very concrete realisation of the asymptotic safety ideas. The RG flow also has a potential to explain why the dimensionless parameter in the Lagrangian of GR (\ref{f-GR}) is so large -- which is a rephrasal of the infamous cosmological constant problem. 

While these possibilities are intriguing, many questions remain open and much more work is needed to see if the ideas described here can be realised. First and foremost it is necessary to study how (\ref{action}) gets renormalised in the quantum theory, and see if the expectation of closedness under the renormalisation holds. This can be done via the standard textbook methods at least at one-loop level. The required Feynman rules will soon appear in \cite{DKS}. Alternatively, one can use the more sophisticated technology of the background field method and the the heat kernel expansion. Work on both of these lines of investigation is in progress. Once it is seen that the class (\ref{action}) is preserved by the renormalisation, the RG flow will be calculable and the ideas of asymptotic safety can be probed. 

Another very important line of research is that of finding which types of matter can naturally be coupled to gravity in this formalism. As we have discussed above, the only natural coupling in our approach appears to be by enlarging the gauge group from ${\rm SU}(2)$ that desribes pure gravity to a general group $G$. It would be very interesting to see what type of matter can appear this way. The bosonic case has already been studied in \cite{Krasnov:2011hi} and, in particular, it was found that massless gauge bosons and massive fields of various spins can be described. An interesting version of the spontaneous (gauge) symmetry breaking scenario was also found in this context. It would be extremely interesting to see if the usual Dirac and Weyl fermions can be coupled to gravity by enlarging the gravitational ${\rm SU}(2)$ to some super-group.  

Another important open problem is to understand physical implications of the parity-violating character of the present class of theories. The question is already interesting in the context of pure gravity, for it can be e.g. shown that the propagation of the gravitational waves on any time-dependent (but not conformally flat) homogeneous isotropic background is left-right asymetric in a general theory from the class (\ref{action}). Since it is easy to imagine that departures from GR described by (\ref{action}) could be present in the very early Universe, it is interesting to determine whether there can be any observable effects of the parity asymetry. This set of questions is also currently under the investigation. 

\section*{Acknowledgments} The author was partially supported by a fellowship from the Alexander von Humboldt Foundation, Germany, and by the ERC Starting Independent Researcher Grant 277570-DIGT. The hospitality of the Albert Einstein Institute, Golm, during the period while this paper was written is gratefully acknowledged. The author is also grateful to the Proceedings of the Royal Society Editor that has invited this review, for patience with him in particular.

\end{document}